\documentclass[manuscript,screen]{acmart}

\AtBeginDocument{%
  }

\setcopyright{acmlicensed}
\copyrightyear{2018}
\acmYear{2018}
\acmDOI{XXXXXXX.XXXXXXX}

\acmConference[Conference acronym 'XX]{Make sure to enter the correct
  conference title from your rights confirmation emai}{June 03--05,
  2018}{Woodstock, NY}
\acmISBN{978-1-4503-XXXX-X/18/06}

\usepackage{booktabs}
\usepackage{multirow}
\usepackage{array}
\usepackage{multirow}
\usepackage{boldline}
\usepackage{makecell}
\usepackage{multirow}
\usepackage{tabularx}   

\begin{document}

\title{Embedding in Recommender Systems: A Survey\\}


\author{Maolin Wang}
\authornote{Maolin Wang, Xinjian Zhao, and Wanyu Wang contributed equally to this paper.}
\email{maolin.wang@cityu.edu.hk}
\affiliation{%
  \institution{City University of Hong Kong}
  \city{Hong Kong}
  \country{China}
}

\author{Xinjian Zhao}
\authornotemark[1]
\email{xinjian.zhao@cityu.edu.hk}
\affiliation{%
  \institution{City University of Hong Kong}
  \city{Hong Kong}
  \country{China}
}

\author{Wanyu Wang}
\authornotemark[1]
\email{wanyuwang4-c@my.cityu.edu.hk}
\affiliation{%
  \institution{City University of Hong Kong}
  \city{Hong Kong}
  \country{China}
}
\author{Sheng Zhang}
\email{szhang844-c@my.cityu.edu.hk}
\affiliation{%
  \institution{City University of Hong Kong}
  \city{Hong Kong}
  \country{China}
}

\author{Jiansheng Li}
\email{jsli@cityu.edu.hk}
\affiliation{%
  \institution{City University of Hong Kong}
  \city{Hong Kong}
  \country{China}
}

\author{Bowen Yu}
\email{bowyu2-c@my.cityu.edu.hk}
\affiliation{%
  \institution{City University of Hong Kong}
  \city{Hong Kong}
  \country{China}
}

\author{Binhao Wang}
\email{binhawang2-c@my.cityu.edu.hk}
\affiliation{%
  \institution{City University of Hong Kong}
  \city{Hong Kong}
  \country{China}
}

\author{Shucheng Zhou}
\email{sczhou@cityu.edu.hk}
\affiliation{%
  \institution{City University of Hong Kong}
  \city{Hong Kong}
  \country{China}
}

\author{Dawei Yin}
\email{dawei.yin@baidu.com}
\affiliation{%
  \institution{Baidu Inc.}
  \city{Beijing}
  \country{China}
}

\author{Qing Li}
\email{qing.li@polyu.edu.hk}
\affiliation{%
  \institution{Hong Kong Polytechnic University}
  \city{Hong Kong}
  \country{China}
}

\author{Ruocheng Guo}
\email{rguo.asu@gmail.com}
\affiliation{%
  \institution{Unaffiliated Researcher}
  \city{Hong Kong}
  \country{China}
}

\author{Xiangyu Zhao}
\authornote{The corresponding author}
\email{xianzhao@cityu.edu.hk}
\affiliation{%
  \institution{City University of Hong Kong}
  \city{Hong Kong}
  \country{China}
}
\renewcommand{\shortauthors}{Maolin Wang et al.}

\begin{abstract}
Recommender systems have become an essential component of many online platforms, providing personalized recommendations to users. 
A crucial aspect is embedding techniques that convert the high-dimensional discrete features, such as user and item IDs, into low-dimensional continuous vectors, which can enhance the recommendation performance. 
Embedding techniques have revolutionized the capture of complex entity relationships, generating significant research interest. This survey presents a comprehensive analysis of recent advances in recommender system embedding techniques. We examine centralized embedding approaches across matrix, sequential, and graph structures.
In matrix-based scenarios, collaborative filtering generates embeddings that effectively model user-item preferences, particularly in sparse data environments. For sequential data, we explore various approaches including recurrent neural networks and self-supervised methods such as contrastive and generative learning. In graph-structured contexts, we analyze techniques like node2vec that leverage network relationships, along with applicable self-supervised methods.
Our survey addresses critical scalability challenges in embedding methods and explores innovative directions in recommender systems. We introduce emerging approaches, including AutoML, hashing techniques, and quantization methods, to enhance performance while reducing computational complexity. Additionally, we examine the promising role of Large Language Models (LLMs) in embedding enhancement.
Through detailed discussion of various architectures and methodologies, this survey aims to provide a thorough overview of state-of-the-art embedding techniques in recommender systems, while highlighting key challenges and future research directions.
To facilitate development, evaluation, and comparison of embedding-based recommender systems, we provide an open-source repository~\footnote{https://github.com/Applied-Machine-Learning-Lab/Embedding-in-Recommender-Systems}.
 \end{abstract}





\maketitle

\section{Introduction}

Recommender systems have become an integral part of online platforms, providing personalized suggestions to users based on their preferences and behavior~\cite{zangerle2022evaluating, wang2025star, wang2025findrec}. One of the key components of a recommender system is the representation of {high-dimensional discrete features}
in a {low-dimensional continuous vector} space, known as embeddings.
Embeddings have been shown to improve the performance of recommender systems by capturing the complex relationships between entities (e.g., items and users).
Although embeddings have been widely employed in recommender systems, there has been a burgeoning body of research dedicated to the investigation of them. Various methodologies have been proposed to generate entity embeddings, with the aim of capturing intricate relationships between entities.
These methods include collaborative filtering~\cite{rendle2010factorization,abdi2018matrix,chen2022survey}, self-supervised learning~\cite{yu2022SelfSurvey}, and graph-based techniques~\cite{gao2021graph}. 
Additionally, Auto Machine Learning~(AutoML)~\cite{liu2020unas}, hashing techniques~\cite{kang2020learning,ghaemmaghami2022learning}, and quantization techniques~\cite{lu2010survey,ramanan2012review} have also been explored as ways to improve the performance and efficiency of embedding-based recommender systems.

Collaborative filtering (CF) is a widely adopted technique in recommender systems~\cite{koren2021advances} for matrix data format. This method hinges on generating embeddings, which are compact, low-dimensional representations of users and items. These embeddings serve as a means to capture the underlying preferences and intrinsic characteristics of users and items alike. Within collaborative filtering, two prominent strategies for generating embeddings emerge: matrix factorization and factorization machines~\cite{rendle2010factorization}. The allure of CF lies in its ability to provide accurate recommendations by discerning intricate relationships and interactions among users and items. It notably excels in handling sparse data and effectively tackling the challenge of the cold-start problem, where limited information about new items or users is available~\cite{koren2021advances}. 
However, generating embeddings through collaborative filtering (CF) requires extensive training on large datasets, potentially imposing a significant computational burden.
Moreover, it is worth noting that CF approaches might exhibit suboptimal performance when dealing with data that is inherently limited or predisposed to biases.

In response to some of the limitations traditional CF poses, self-supervised learning-based techniques have arisen as a promising solution for embedding the learning of recommender systems~\cite{huang2022self} {with limited data}. 
These methodologies commonly harness the power of contrastive learning or generative learning techniques to derive entity embeddings. By capitalizing on expansive datasets and tapping into non-linear relationships, these methods aptly capture the latent semantics embedded within the data~\cite{wu2021SGL}. This equips them with remarkable generalizability across various tasks and domains.
Yet, the success of self-supervised learning pivots significantly on the meticulous crafting of data augmentation methodologies, alongside the judicious selection of model architectures, loss functions, and pre-trained models. These factors wield considerable influence over the stability and convergence of the training process. It is important to note that self-supervised learning places particular emphasis on the graph and sequence structures of data, which will be a central focus of this survey.

Structure-based methods, such as node2vec~\cite{grover2016node2vec} and graph autoencoders~\cite{simonovsky2018graphvae, ding2021semi} in graph modeling, and sequence modeling techniques like LSTM, GRU, and Transformer, harness rich structural intricacies to create embeddings that capture inherent connections. By effectively capturing the multifaceted network patterns and associations present in recommendation scenarios, these methods yield embeddings that encapsulate the intricate relationships between entities. These techniques are particularly significant in settings with complex underlying structures, such as social network environments and sequential recommendation scenarios like e-commerce browsing history and video watching sequences. Furthermore, they provide a means to incorporate auxiliary information through item-item and user-user relationships~\cite{fan2019graph}.

One of the main challenges in the field of embeddings in recommender systems is the scalability of the methods. As the number of items and users in a recommender system becomes massive, the computational cost of generating embeddings can be a concern. 
To enhance efficiency and reduce complexity, various research directions have been explored for embedding-based recommender systems.
At the forefront of this evolution within recommendation systems stand auto machine learning, hashing, and quantization techniques.
Auto machine learning (AutoML) automates optimal machine learning solutions, streamlining tasks like model architecture design and hyperparameter tuning. In recommendation systems, AutoML aids in selecting suitable embedding sizes, boosting performance while mitigating challenges~\cite{zhaok2021autoemb}.
Hashing techniques~\cite{weinberger2009feature} address the sparsity challenges of one-hot encodings by leveraging hashing functions for dimension reduction, minimizing storage, and computation. 
Quantization compresses high-dimensional embeddings into compact codes, curbing memory demands and enhancing efficiency. This process reconstructs original embeddings with minimal distortion, offering an efficient solution~\cite{jegou2010product,ge2013optimized,lian2020product}.

Overall, this survey seeks to offer a thorough panoramic view of embedding applications for recommender systems. We will delve into three categories of widely utilized techniques for embedding generation, encompassing (1) collaborative filtering, (2) self-supervised learning, and (3) graph-based approaches. Furthermore, we will scrutinize integrating (a) AutoML, (b) hashing techniques, and (c) quantization methodologies into the embedding framework. Moreover, in each chapter, we demonstrate existing surveys and unveil prospective pathways for future advancement. Finally, we will briefly showcase how embeddings are practically applied in Click-Through-Rate~(CTR) predictions and Large Language Models~(LLMs) based recommendations, providing a practical outlook on the embedding usage. To the best of our knowledge, this paper represents the first systematic review of embedding techniques in recommendation systems.
\section{Embedding Learning of Matrix Format}
\label{sec:matrix_format}
In this section, we will explore traditional embedding methods in recommender systems, specifically Collaborative Filtering (CF)~\cite{sarwar2001item}.
CF primarily focuses on learning the similarities between users and items as well as user behavior to predict potential item preferences for recommendations, commonly employing matrix format modeling.
The two typical methods used to implement embeddings in CF are Matrix Factorization (MF) and Factorization Machines (FM).
Matrix Factorization~\cite{koren2008factorization} involves decomposing the user rating matrix into a generalized product of the user feature matrix and the item feature matrix, thereby facilitating recommendations based on user preferences for items.
These preferences are calculated using the corresponding user and item embedding vectors.
However, simple MF methods face challenges with the sparsity of the rating matrix due to the extensive use of one-hot encoded categorical features, and they cannot utilize auxiliary features about users and items.
To address these issues, FM~\cite{rendle2010factorization} is introduced as an initial learning strategy that enhances prediction performance by learning the similarities between feature embeddings derived from one-hot encoding.
By emphasizing matrix formats, this method not only highlights its mathematical structure but also its pivotal role in embedding learning, especially in handling large-scale sparse data. A summary of the methods discussed in this section is presented in Table~\ref{tab:matrix_format_final_complete_with_ref}.
Generally, CF aims to learn the similarity between users and items as well as user behavior, then predict potential preferred items for recommendation.

Then, MF implements machine learning models to fill the missing ratings of the original user-item matrix for the final recommendation task.

\subsection{Matrix Factorization Scheme}
\label{ssec:mf_scheme}

\begin{table}[t!]
\centering
\caption{Embedding Learning of Matrix Format}
\label{tab:matrix_format_final_complete_with_ref}
\begin{tabular}{>{\raggedright}m{2.5cm}|>{\raggedright}p{5cm}|m{6.5cm}}
\toprule
\multirow{6}{2.5cm}{\raisebox{-3\height}{\textbf{\begin{tabular}{@{}l@{}}Embedding \\ Learning of \\ Matrix Format\end{tabular}}}} & 
\multirow{3}{*}{\raisebox{-4\height}{\textbf{Matrix Factorization Scheme (\S\ref{ssec:mf_scheme})}}} & 
\textbf{SVD-based Enhancements:} FunkSVD~\cite{funk2006netflix}, NSVD~\cite{paterek2007improving}, SVD++~\cite{koren2008factorization}, SVDfeature~\cite{chen2012svdfeature}, DELF~\cite{cheng2018delf}, BiasSVD~\cite{koren2009matrix}, TimeSVD~\cite{koren2009collaborative} \\
& & \textbf{Item Similarity Models:} SLIM~\cite{ning2011slim}, FISM~\cite{kabbur2013fism}, NAIS~\cite{he2018nais}, SGNS~\cite{levy2014neural} \\
& & \textbf{Content-aware Models:} ConvMF~\cite{kim2016convolutional}, ConvSeq-MF~\cite{khan2025convseq} \\
\cmidrule(l){2-3}
& 
\multirow{3}{*}{\raisebox{-3\height}{\textbf{Factorization Machines (\S\ref{ssec:fm})}}} & 
\textbf{Foundation Model:} Factorization Machines (FM)~\cite{rendle2010factorization} \\
& & \textbf{Deep Learning Hybrids:} FNN~\cite{zhang2016deep}, PNN~\cite{qu2016product}, NeuFM~\cite{he2017neural}, AFM~\cite{xiao2017attentional} \\
& & \textbf{Wide \& Deep Architectures:} Wide \& Deep~\cite{cheng2016wide}, DeepFM~\cite{guo2017deepfm}, DCN~\cite{wang2017deep}, xDeepFM~\cite{lian2018xdeepfm} \\
\bottomrule
\end{tabular}
\end{table}

MF aims to connect, represent the users and items according to the latent feature factorized.
The singular value decomposition (SVD) is the most common kind of method for matrix factorization, for which the SVD is introduced as a pseudo form in FunkSVD model~\cite{funk2006netflix}.
Intuitively, it involves an approximate factorization of the original rating matrix $\mathbf{R}\in\mathbb{R}^{{m \times n}}$ ($m$ users and $n$ items) to obtain smaller user matrix $\mathbf{U}\in\mathbb{R}^{{m \times d}}$ and item matrix $\mathbf{V}\in\mathbb{R}^{{n \times d}}$, with a hidden embedding dimension $d$:
\begin{equation}
    \begin{aligned}
        \mathbf{R} \approx \mathbf{U} \times \mathbf{V}^T.
    \end{aligned}
\end{equation}
\begin{figure}[htbp]
\centering
\includegraphics[scale=0.6]{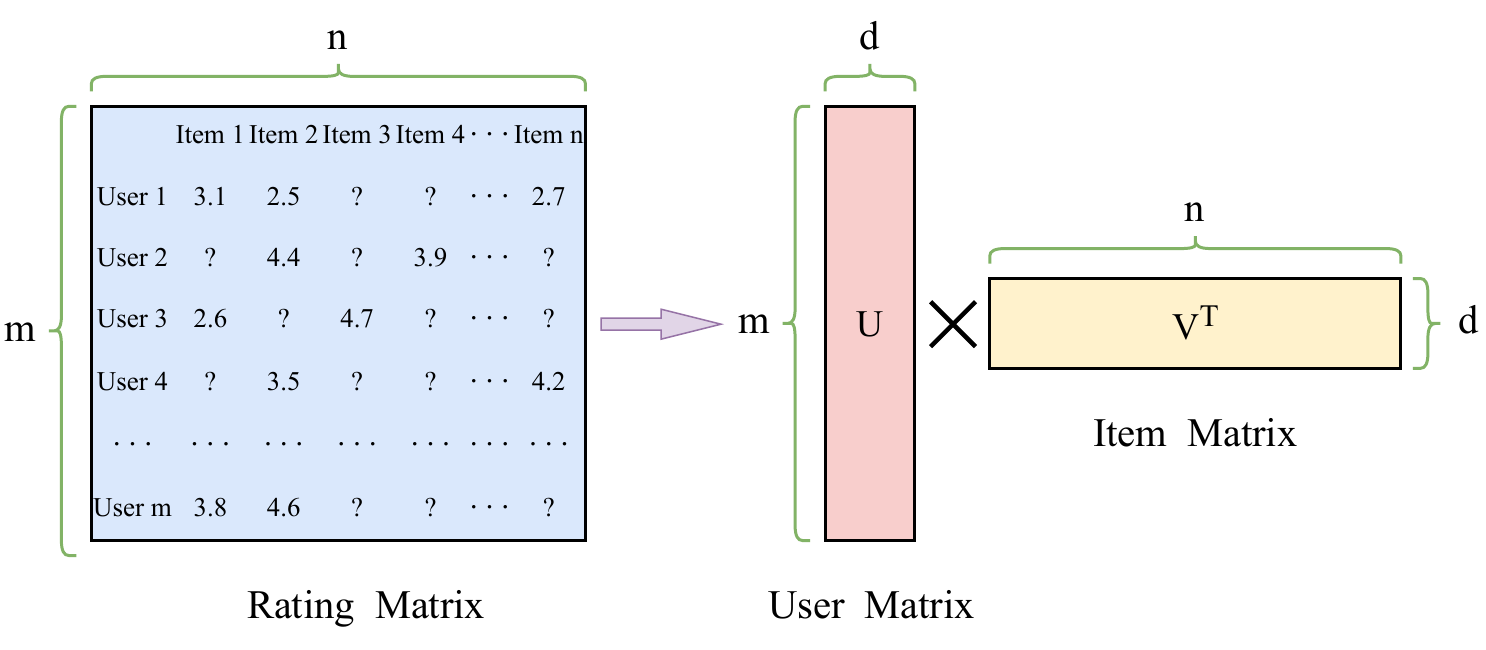}
\captionsetup{justification=raggedright,singlelinecheck=false}
\caption{FunkSVD model~\cite{funk2006netflix}. It involves an approximate factorization on matrix $\mathbf{R}\in\mathbb{R}^{{m \times n}}$ to obtain smaller user matrix $\mathbf{U}\in\mathbb{R}^{{m \times d}}$ and item matrix $\mathbf{V}\in\mathbb{R}^{{n \times d}}$, with a hidden embedding dimension $d$.}
\label{figure}
\vspace{-3mm}
\end{figure}
$\mathbf{U}$ and $\mathbf{V}$ represent unexplainable latent features of the users and the items, respectively.
The final optimization object is
\begin{equation}
    \begin{aligned}
        \mathop{\rm{min}}\limits_{\mathbf{u}_i,\mathbf{v}_j} \sum_{\{(i,j)|\mathbf{R}_{ij} \neq \phi\}} (\mathbf{R}_{ij}-\mathbf{v}_j^T \mathbf{u}_i)^2+\lambda(\left\|\mathbf{u}_i\right\|^2+\left\|\mathbf{v}_j\right\|^2),
    \end{aligned}
\end{equation}
where $\mathbf{u}_i$ and $\mathbf{v}_j$ are the latent feature vectors of user $i$ and item $j$ in d-dimensional space respectively.
$\mathbf{u}_i$ and $\mathbf{v}_j$ are aslo $i$-th and $j$-th column vector of $\mathbf{U}$ and $\mathbf{V}$ respectively.
And the corresponding rating is predicted as $\mathbf{u}_i^T\mathbf{v}_j$,

Since then, many methods with improvements and variations based on FunkSVD~\cite{funk2006netflix} have been gradually presented.
For user factor reduction, NSVD~\cite{paterek2007improving} combines the item factors in a linear way to represent users.
SVD++~\cite{koren2008factorization} then replaces the representing step with feeding latent user factors to optimize NSVD~\cite{paterek2007improving} by rating through a sparse weight matrix.
Then, SVDfeature~\cite{chen2012svdfeature} was presented as a toolkit that combines a rank model.
DELF~\cite{cheng2018delf} implements the attention~\cite{vaswani2017attention} to automatically differentiate interacted items whose importance is different.
BiasSVD~\cite{koren2009matrix} adds a bias part: user-independent or item-independent factors, to FunkSVD~\cite{funk2006netflix}.
TimeSVD~\cite{koren2009collaborative} adds time weights by learning a parameter in each period and training with related data.
While these traditional matrix factorization variants have established strong foundations, recent research has identified new challenges and opportunities in practical deployment.
Modern applications demand not only accuracy but also security, robustness, and enhanced representation learning capabilities~\cite{nguyen2024manipulating,wu2025graphfm}.
Addressing these emerging needs, Shams et al.~\cite{shams2025attack} tackle the critical issue of security in production recommender systems, introducing attack detection mechanisms that monitor item vector shifts to identify malicious manipulations.
This work highlights the gap between theoretical matrix factorization models and real-world deployment requirements.


{Beyond direct modification of the FunkSVD framework, several frameworks have also been proposed through SLIM~\cite{ning2011slim}.}
SLIM~\cite{ning2011slim} introduces a user behavior matrix whose values $r_{ui}$ are binary: 1 for rated item $i$ by user $u$, 0 otherwise. 
Then replace one of the factorized small matrices. 
But SLIM~\cite{ning2011slim} lacks the ability to compare the similarity of either 
Two items were rated by one user at the same time.
FISM~\cite{kabbur2013fism} solves this by learning the low-dimensional latent vector space without the detailed rating information. 
But it considers that all the items rated by a user are equal for further prediction.
NAIS~\cite{he2018nais} uses attention to select items with more important interactions automatically, which fixes this issue well.
SGNS~\cite{levy2014neural} implements skip-gram~\cite{mikolov2013distributed} and negative sampling techniques to factorize a shifted PMI matrix.
ConvMF~\cite{kim2016convolutional} integrates the CNN into PMF~\cite{mnih2007probabilistic} and captures contextual information about the document to improve the performance of rating prediction.
Building upon ConvMF's integration of textual content, the evolution toward comprehensive multimodal matrix factorization continues to gain momentum.
Khan et al.~\cite{khan2025convseq} advance this direction with ConvSeq-MF, which systematically combines convolutional neural networks with sequence modeling while leveraging both user and item auxiliary content information.
Furthermore, the integration of large language models into matrix factorization has opened new possibilities for semantic understanding.
Liu et al.~\cite{liu2025understanding} demonstrate how LLMs can enhance traditional collaborative filtering by providing semantic aspect-aware review exploitation, bridging the gap between textual content and latent factor models.
\subsection{Factorization Machines}
\label{ssec:fm}
A solution to address classification challenges with large-scale sparse data lies in feature combination.
Factorization Machines (FM) \cite{rendle2010factorization} effectively compress numerous sparse features, reduce parameter magnitude, and maintain performance simultaneously.
Fundamentally, FM substitutes the interaction term in regular logistic regression with the dot product of latent factors.
A strategy involves concatenating one-hot vectors representing user and item IDs to create the input feature vector $\mathbf{x}$.
Alternatively, auxiliary feature vectors linked to users and items can also be combined. Second-order feature combinations are then considered.
For input feature vector $\mathbf{x}$, the output value expression $\Phi_{\rm FM}(\mathbf{x})$ of degree $d=2$ in factorization machines is given by:

\begin{equation}
    \begin{aligned}
        \Phi_{\rm FM}(\mathbf{x}) = w_0 + \sum ^n_{i=1}w_ix_i + \sum ^n_{i=1} \sum ^n_{j=i+1} \left< \mathbf{v}_i,\mathbf{v}_j \right> x_ix_j,
    \end{aligned}
\end{equation}
where computation time is linear due to the concept of introducing high-order interactions based on linear regression.
The dot product of $k$-dimensional vectors $\mathbf{v}_i$ and $\mathbf{v}_j$ is:

\begin{equation}
    \begin{aligned}
        \left< \mathbf{v}_i,\mathbf{v}_j \right> = \sum ^k_{f=1} \mathbf{V}_{i,f} \mathbf{V}_{j,f}.
    \end{aligned}
\end{equation}
Parameters $\mathbf{w} = (w_1,\cdots,w_n)^T \in \mathbb{R}^{n}$ and $\mathbf{V} = (\mathbf{v}_1,\cdots,\mathbf{v}_k)^T\in\mathbb{R}^{n\times k}$ are used, where $w_0$ and $w_i$ represent bias and parameter for the $i$-th variable respectively.
$k$ signifies the dimension of learned latent embeddings, while $n$ denotes the dimension of the original feature vector.
FM facilitates model resolution by replacing $w_{i,j}$ with the dot product of vectors $v_i$ and $v_j$, effectively achieving the factorization process.
When FM exclusively uses user and item IDs with binary interaction outputs, its prediction process aligns with FunkSVD \cite{DBLP:journals/corr/abs-2203-11026}.
Nonetheless, FM extends its reach by incorporating the embedding mechanism's features and extending to diverse feature sets, bolstering flexibility and applicability.
While factorization machines effectively handle sparse features through second-order interactions, they face computational challenges when dealing with numerous categorical fields in real-time scenarios.
Recent work has explored more efficient approaches for handling field interactions in latency-critical applications.
Low-rank field-weighted factorization machines~\cite{shtoff2024low} explore the low-rank decomposition of field interaction matrices to reduce computational complexity while maintaining modeling expressiveness for online serving scenarios.
Moreover, the limitation of modeling only pairwise user-item interactions has been highlighted, as real-world systems often exhibit complex, higher-order feature dependencies.
To address this, GraphFM~\cite{wu2025graphfm} introduces a graph-based factorization machine framework that explicitly models higher-order feature interactions via graph neural networks, significantly extending the expressive power of traditional FM models.

Deep Neural Networks (DNNs) excel in capturing high-order feature interactions \cite{rendle2010factorization}. Factorization Neural Networks (FNNs) \cite{zhang2016deep} enhance this capability by integrating DNNs as the lower layer, enabling the learning of complex interactions that FM's linear approach lacks.
FNN \cite{zhang2016deep} reduces dimensionality by transforming sparse features using supervised learning embedding layers.
It reduces the need for feature engineering, enhances model expressiveness, and offers inspiration for FM-MLP combinations.
Product-based Neural Networks (PNN) \cite{qu2016product} introduces a product layer comprising linear and nonlinear segments to explicitly calculate the product expression of 2-dimensional feature interactions, while it may perform well with MLPs like FNN \cite{zhang2016deep}.
%
Neural Factorization Machines (NeuFM) \cite{he2017neural} follow a similar linear and nonlinear composition, fusing 2-dimensional linear features (derived from a general matrix factorization layer) with high-dimensional nonlinear features from DNNs.
Attentional Factorization Machines (AFM) \cite{xiao2017attentional} introduce an attention mechanism, learning an attention score for feature interaction importance through a neural network.
This mechanism reflects varying feature importance, enhancing model sensitivity.

Contrasting the aforementioned method, a fusion of factorization machines (FM) and deep neural networks (DNNs) provides an avenue for alleviating the manual feature engineering burden inherent in certain DNNs.
Notably, Google's Wide \& Deep~\cite{cheng2016wide} represents a significant advancement in recommendation systems.
This model effectively tackles the challenges of memorization and generalization by capturing interactions between low-dimensional and high-dimensional features.
However, it still requires artificial feature engineering for the Wide component's input.
Huawei's DeepFM~\cite{guo2017deepfm} builds upon the Wide \& Deep framework, replacing the Wide component with FM.
By sharing the embedding part between FM and Deep components, training efficiency improves.
Importantly, DeepFM dispenses with pre-training FM and artificial feature engineering, achieving simultaneous learning of low-dimensional and high-dimensional feature interactions.
Further progress, building upon Wide \& Deep~\cite{cheng2016wide} and DeepFM~\cite{guo2017deepfm}, centers on enhancing explicit higher-dimensional feature interactions, particularly through cross-product computations.
Deep \& Cross Network (DCN)~\cite{wang2017deep}, an initial approach in this direction, adopts a cross-network instead of FM, replacing the Wide component.
While promising, DCN might not universally suit all scenarios.
xDeepFM~\cite{lian2018xdeepfm} achieves simplification by presenting a compressed interaction network (CIN) based on DCN~\cite{wang2017deep}, and the way of learning feature interaction is changed to a vector-wise way.
However, AutoInt differentiates itself by sidestepping the combination of a cross-network and a deep neural network.
Instead, it directly integrates an interacting layer after the embedding phase.

\subsection{Surveys and Future Directions}
\par \noindent \textbf{Survey.}
Abdi et al.\cite{abdi2018matrix} focus on the field of context-aware recommendation based on matrix factorization, while Chen et al.~\cite{chen2022survey} analyze and conclude the theory of nonnegative matrix factorization with DNN.
LibFFM\footnote{LibFFM: https://github.com/ycjuan/libffm} and xlearn\footnote{xlearn: https://github.com/aksnzhy/xlearn} are used to achieve FFM~\cite{juan2016field}. For other algorithms, corresponding toolkits can also be found.
Such as DeepCTR\footnote{DeepCTR: https://github.com/shenweichen/DeepCTR}, which can achieve various CTR prediction models, including DeepFM~\cite{guo2017deepfm}.

\noindent \textbf{Future Direction: Synergistic Model Fusion for Complex Scenarios.}{
An interesting avenue for future exploration lies in the synergy between different models.
Instances like CAT-E and CAT-NT, as proposed by Zhao et al.~\cite{zhao2017gb}, creatively combine SVD, FM, and gradient boosting, showcasing the potential of model fusion.
These approaches cater to highly specialized scenarios, often tied to the designs of MF models based on SVD and FM.
Investigating and expanding upon such combinations could yield even more tailored solutions for intricate use cases.}
Except for the MF based on SVD \& LFM, and FM, various kinds of CF development are also very meaningful for figuring out different specific scenarios in specific applications.
For instance, DMF~\cite{xue2017deep} uses explicit rating and non-preference implicit feedback to map the information of users and items to a low-dimensional space nonlinearly, which combines both the explicit and implicit information.
EMF~\cite{abdollahi2016explainable} combines the CF and graph weight, and XPL-CF~\cite{almutairi2021xpl} proposes to make the user-item relation clear.

\makeatletter
\renewcommand{\maketag@@@}[1]{\hbox{\m@th\normalsize\normalfont#1}}%
\makeatother
\section{Sequential Data Embedding}
\label{sec:seq_embedding}
Sequential data, characterized by ordered interactions (e.g., user-item interaction sequences), is central to many recommender systems.
Learning effective embeddings for such sequences requires capturing both temporal dependencies and higher-order patterns.
Sequence embedding methods have evolved from RNN and CNN-based approaches to Transformer and state-space architectures, often enhanced with self-supervised learning (SSL) techniques. A detailed breakdown of these methods is provided in Table~\ref{tab:sequential_embedding_final}.
\begin{table}[t!]
\centering
\caption{Sequential Data Embedding}
\label{tab:sequential_embedding_final}
\begin{tabular}{>{\raggedright}m{2.5cm}|>{\raggedright}p{5.5cm}|m{6cm}}
\toprule
\multirow{8}{2.5cm}{\raisebox{-4\height}{\textbf{\begin{tabular}{@{}l@{}}Sequential \\ Data \\ Embedding\end{tabular}}}} & 
\multirow{3}{*}{\raisebox{-4\height}{\textbf{Sequential Modeling Techniques (\S\ref{ssec:seq_modeling})}}} & 
\textbf{RNN/CNN-based Models:} GRU4Rec~\cite{hidasi2015session}, NARM~\cite{li2017neural}, Caser~\cite{tang2018personalized}, GLINT-RU ~\cite{zhang2025glint} \\
& & \textbf{Transformer-based Models:} SASRec~\cite{kang2018self}, BERT4Rec~\cite{sun2019bert4rec} \\
& & \textbf{Graph \& State-Space Models:} SR-GNN~\cite{wu2019session}, Mamba4rec~\cite{liu2024mamba4rec}, STAR-Rec~\cite{wang2025star}, DiffuRec~\cite{li2023diffurec} \\
\cmidrule(l){2-3}
& 
\textbf{Contrastive Learning for Sequential Data Embeddings (\S\ref{ssec:contrastive_learning})} & 
\textbf{Key Models:} CL4SRec~\cite{xie2022CL4Rec}, CoSeRec~\cite{liu2021CoSeRec}, ContraRec~\cite{wang2022ContraRec} \\
\cmidrule(l){2-3}
& 
\multirow{3}{*}{\raisebox{-2\height}{\textbf{\begin{tabular}{@{}l@{}}Generative Self-supervised Learning \\ Methods (\S\ref{ssec:generative_ssl})\end{tabular}}}} & 
\textbf{Foundation Model:} BERT4Rec~\cite{sun2019bert4rec, devlin2018bert} \\
& & \textbf{Model Extensions:} UNBERT~\cite{zhang2021unbert}, U-BERT~\cite{qiu2021ubert}, UPRec~\cite{xiao2021uprec} \\
& & \textbf{General Embedding Models:} PeterRec~\cite{yuan2020PeterRec}, ShopperBERT~\cite{shin2021shopperbert} \\
\bottomrule
\end{tabular}
\end{table}
\subsection{Sequential Modeling Techniques}
\label{ssec:seq_modeling}
Early sequence modeling in recommendation systems relied on recurrent or convolutional architectures.
RNN-based models such as GRU4Rec~\cite{hidasi2015session} and NARM~\cite{li2017neural} encode sequences into hidden embeddings to capture temporal dependencies, while CNN-based models like Caser~\cite{tang2018personalized} use convolutional filters to capture local sequential patterns efficiently.
Transformer-based models, including SASRec~\cite{kang2018self} and BERT4Rec~\cite{sun2019bert4rec}, leverage self-attention to model long-term semantics, enhancing representation learning.
Graph-based models such as SR-GNN~\cite{wu2019session} construct graphs over sequence elements to capture higher-order transitions, and recent state-space or diffusion models (e.g., Mamba4rec~\cite{liu2024mamba4rec}, DiffuRec~\cite{li2023diffurec}) efficiently model ultra-long or uncertain sequences.
These architectures form the backbone of modern sequential recommendation, providing diverse ways to encode sequential dependencies and latent patterns in user-item interactions.

\subsection{Contrastive Learning for Sequential Data Embeddings}
\label{ssec:contrastive_learning}
\begin{figure}[htbp]
\centering
\includegraphics[scale=0.8]{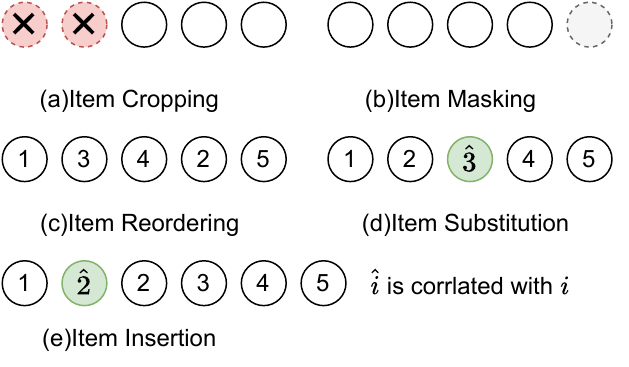}
\centering
\caption{Augmentation methods of Sequence. (a) Crop items for contrastive learning. (b) Mask item elements to prevent overfitting. (c) Augment data by reordering items, enhancing recommenders on varying sequences. (d) Substitute highly correlated yet possibly redundant items. (e) Add related items before the original item.}
\vspace{-3mm}
\label{seqAug}
\end{figure}

For sequence data, this section introduces four primary techniques to generate corrupted sequences to facilitate sequence augmentation.
\textbf{Item Cropping}, inspired by image cropping in computer vision, is applied to contrastive learning in sequential recommendation.
\textbf{Item Masking}, borrowed from language models, involves masking elements in a sequence to prevent overfitting and encourage learning of high-order structures among tokens.
Similarly, \textbf{Item Reordering} is proposed to ensure predictions remain invariant to item order, training recommendation models on augmented data with varying item sequences.
\textbf{Item Substitution} enhances recommendation models by maintaining the original sequence's structure while substituting highly correlated yet possibly redundant items.
Item Insertion, akin to substitution, involves adding related items before each original item in a subset of the sequence.
Specific details are shown in Fig.~\ref{seqAug}.

Following the general pipeline of CL, the CL for sequences first generates different views of each sequence by the augmentation methods mentioned above.
Then train the recommendation model with contrastive loss (e.g., SimCLR~\cite{chen2020simCLR}). {CL4SRec}~\cite{xie2022CL4Rec} deals with the sequential recommendation task.
It adopts three augmentation methods: item cropping, masking, and reordering.
Given a minibatch of $N$ users, it generate 2$N$ augmented sequences $\{S_{u_{1}}^{a_{i}}, S_{u_{1}}^{a_{j}}, S_{u_{2}}^{a_{i}}, S_{u_{2}}^{a_{j}}, \ldots, S_{u_{N}}^{a_{i}}, S_{u_{N}}^{a_{j}}\}$, where $S_{u_{n}}^{a_{i}}$ denotes the sequence interacted by user $u_{n}$ augmented by the augmentation method $a_{i}$ (e.g., item cropping).
Then, for each user $u_{i}$, it regards $\left(S_{u_{i}}^{a_{i}}, S_{u_i}^{a_{j}}\right)$ as the positive pair, and other $2(N-1)$ elements as negative samples.
Additionally, it utilizes the transformer encoder to compute the sequence embeddings.
Eventually, the softmax cross-entropy loss is used to distinguish the positive pairs from the negative ones.
{Similar to CL4SRec, {CoSeRec} \cite{liu2021CoSeRec} uses item substitution and insertion for data augmentation, while {ContraRec} \cite{wang2022ContraRec} defines positive pairs from both the same sequence and different sequences with the same target item.} 

{Contrastive methods utilize the CL loss with augmented data to enhance embeddings by capturing latent patterns.
However, selecting augmentation techniques is difficult, as there's no fixed rule for their choice.
This makes it necessary for researchers to choose data augmentation strategies carefully.
While contrastive methods effectively capture invariant patterns through augmentation, they face challenges in selecting appropriate transformation strategies.
This limitation has led researchers to explore complementary approaches, particularly generative methods that learn through reconstruction rather than discrimination.
} 

\subsection{Generative Self-supervised Learning Methods}
\label{ssec:generative_ssl}
Generative methods in sequential recommendation leverage the principle of learning through reconstruction, inspired by the success of masked language modeling in NLP \cite{devlin2018bert}.
These approaches corrupt input sequences through masking or other perturbations, then train models to recover the original information, thereby learning robust sequential representations without explicit supervision.
{BERT4Rec} \cite{sun2019bert4rec} utilizes the BERT \cite{devlin2018bert} encoder for sequential recommendation. In particular, it uses BERT to model sequential user behaviors.
    BERT4Rec first applies item masking on the sample sequence, where the masked items are replaced with the special token '[mask]'.
Then it trains the model to predict the original item being masked based on the items nearby in the sequence.
Its loss is defined as:
    \begin{equation}
    \label{bertloss}
        \mathcal{L}=\frac{1}{\left|\mathcal{V}_{u}^{m}\right|} \sum_{v_{m} \in \mathcal{V}_{u}^{m}}-\log P\left(v_{m}=v_{m}^{*} \mid \hat{\mathcal{S}_{u}}\right),
    \end{equation}
    where $\hat{\mathcal{S}_{u}}$ is obtained by applying item masking on the original user behavior sequence $\mathcal{S}_{u}$.
$\mathcal{V}_{u}^{m}$ is the set of masked items. $v_{m}^{*}$ is the original item for the masked item $v_{m}$, and $P(v_m|\hat{\mathcal{S}_{u}})$ is the conditional probability calculated by the decoder of BERT.
In the training phase, we can compute the item embedding through the encoder of BERT.
There are several extensions of BERT4Rec for different tasks. For example, {UNBERT} \cite{zhang2021unbert} extends BERT for news recommendation.
{U-BERT} \cite{qiu2021ubert} aims to learn user embeddings of the target domains by transferring knowledge from content-rich source domains.
{UPRec} \cite{xiao2021uprec} extends the BERT4Rec model to handle heterogeneous information such as user attributes and social networks.
Another line of research focuses on learning general item embeddings that are suitable for various recommendation tasks through generative methods.
    {PeterRec} \cite{yuan2020PeterRec} is to conduct learning-to-learn idea \cite{bertinetto2016learningtolearn} on cross-domain recommendation task.
Dissimilar to traditional cross-domain methods \cite{xie2022CCDR,wang2021PCRec}, which update the parameters of the whole model during the fine-tuning phase, it subtly inserts small neural network modules into the backbone of the original model and only trains the small neural networks to adapt to the downstream tasks.
Like the application of BERT \cite{qiu2021ubert} for learning the general word embedding, {ShopperBERT} \cite{shin2021shopperbert} takes advantage of the rich user behaviors to pre-train the BERT model based on nine auxiliary tasks for the general embedding.
{The experiment shows that it outperforms the model designed for one specific downstream recommendation task and indicates that learning the general embedding is feasible.}

While these methods have demonstrated the effectiveness of basic SSL, a key challenge remains in understanding how SSL effectiveness varies across different recommendation contexts.
Recent work has revealed significant performance variations depending on the specific recommendation scenario and data characteristics~\cite{wei2023multi,ren2024sslrec}.
SSLRec~\cite{ren2024sslrec} provides a comprehensive framework enabling systematic comparison of SSL-enhanced models across five recommendation scenarios(e.g., multi-behavior and knowledge-enhanced recommendation).
This unified evaluation reveals that different recommendation tasks require carefully tailored SSL strategies rather than one-size-fits-all approaches.
MENTOR~\cite{wei2023multi} demonstrates that multimodal recommendation benefits from simultaneous intra-modal and inter-modal contrastive objectives, where different information granularities require distinct contrastive strategies - visual features benefit from patch-level contrast while collaborative signals require user-item level contrast.
The model utilizes the graph attention network \cite{velivckovic2017graphattentionnetwork} to learn the ontology embedding from diagnosis and medical tree-like graphs in order to make medication recommendations.
The former task is to predict the masked codes from the same type (i.e., medication or diagnosis) of ontology embeddings, and the latter is to reconstruct the masked codes from different types of ontology embeddings.
They propose MCNSampling, a novel method to select neighboring nodes using importance scores and use BERT-like patterns for node embedding learning.
It first pre-trains a transformer encoder on the embeddings of the first $K$ neighboring nodes to simulate cold-start scenarios.
Then, a GCN model is pre-trained, incorporating the meta embedding in each convolution step.
It finally maximizes cosine similarity between the predicted embedding and the ground-truth embedding generated by NCF \cite{he2017neuralNCF} using all neighbor nodes.

Although it is suitable for the sequence data naturally, researchers extend the methods to deal with graph embedding learning by transforming the graph data into sequence data.
Thus, people should carefully consider the trade-off between the embedding quality and training cost.
\subsection{Surveys and Future Directions}
\noindent  \textbf{Surveys.} While this section focuses on embedding learning for sequential data, several surveys provide comprehensive overviews of sequential recommendation methods~\cite{fang2020deep,dang2024data,pan2024survey,boka2024survey}.
Pan et al.\cite{pan2024survey} offer an extensive survey on sequential recommendation, introducing a new taxonomy that categorizes various approaches, including ID-based, side information-enhanced, and generative models.
Boka et al.\cite{boka2024survey} explore how sequential recommendation systems utilize interaction history to make more accurate and personalized recommendations.
\noindent \textbf{Future direction 1: The choice of augmentation methods.} 
        Augmenting the data with a suitable and effective view is able to generate a strong self-supervised signal to optimize the model effectively.
However, to search for workable augmentation techniques, scientists usually test many strategies randomly, resulting in the waste of labor force and sub-optimal model performance.
Several works \cite{xiao2020theory1}, \cite{kang2020theory2} have studied the theory of augmentation selection in contrastive learning, but less in the recommendation task.
Thus, principles to guide the choice of augmentation methods in recommendation tasks are waiting for further exploration.
\noindent \textbf{Future direction 2: On-device embedding for recommendation model.} 
     Recommendation models handle extensive user and item data, requiring substantial embedding capacity and memory.
Traditional on-device deployment poses challenges due to device limitations. However, knowledge distillation has shown promise in shrinking model size, as demonstrated in works like \cite{wang2021graph_distillation} and \cite{yu2021tiny}.
Integrating self-supervised learning methods could potentially compensate for performance loss in these compact models, an area with limited exploration.
Thus, it is possible to learn a general-purpose embedding in a large-scale dataset that can serve different recommendation tasks directly or be fine-tuned with less computation cost.
Some works \cite{zhang2020generalpurpose}, \cite{shin2021shopperbert}, \cite{yuan2020PeterRec} have proposed some promising performance models, but they all take the BERT-like architecture.
Therefore, it is a challenge to devise a new architecture for general-purpose embedding.

\begin{table}[t!]
\centering
\caption{Sequential Data Embedding}
\label{tab:sequential_embedding_final}
\begin{tabular}{>{\raggedright}m{2.5cm}|>{\raggedright}p{5.5cm}|m{6cm}}
\toprule
\multirow{8}{2.5cm}{\raisebox{-4\height}{\textbf{\begin{tabular}{@{}l@{}}Sequential \\ Data \\ Embedding\end{tabular}}}} & 
\multirow{3}{*}{\raisebox{-5\height}{\textbf{Sequential Modeling Techniques (\S\ref{ssec:seq_modeling})}}} & 
\textbf{RNN/CNN-based Models:} GRU4Rec~\cite{hidasi2015session}, NARM~\cite{li2017neural}, Caser~\cite{tang2018personalized}, GLINT-RU ~\cite{zhang2025glint} \\
& & \textbf{Transformer-based Models:} SASRec~\cite{kang2018self}, BERT4Rec~\cite{sun2019bert4rec} \\
& & \textbf{Graph \& State-Space Models:} SR-GNN~\cite{wu2019session}, Mamba4rec~\cite{liu2024mamba4rec}, STAR-Rec~\cite{wang2025star}, DiffuRec~\cite{li2023diffurec} \\
\cmidrule(l){2-3}
& 
\textbf{Contrastive Learning for Sequential Data Embeddings (\S\ref{ssec:contrastive_learning})} & 
\textbf{Key Models:} CL4SRec~\cite{xie2022CL4Rec}, CoSeRec~\cite{liu2021CoSeRec}, ContraRec~\cite{wang2022ContraRec} \\
\cmidrule(l){2-3}
& 
\multirow{3}{*}{\raisebox{-2\height}{\textbf{\begin{tabular}{@{}l@{}}Generative Self-supervised Learning \\ Methods (\S\ref{ssec:generative_ssl})\end{tabular}}}} & 
\textbf{Foundation Model:} BERT4Rec~\cite{sun2019bert4rec, devlin2018bert} \\
& & \textbf{Model Extensions:} UNBERT~\cite{zhang2021unbert}, U-BERT~\cite{qiu2021ubert}, UPRec~\cite{xiao2021uprec} \\
& & \textbf{General Embedding Models:} PeterRec~\cite{yuan2020PeterRec}, ShopperBERT~\cite{shin2021shopperbert} \\
\bottomrule
\end{tabular}
\end{table}
\section{Graph based Embedding}
\label{sec:graph_embedding}
Graph-based embeddings have gained significant attention in recommender systems due to their ability to leverage topological information in graph data.
This capability has been greatly enhanced by recent advancements in graph machine learning, which enable models to effectively capture complex relationships and produce high-quality node and graph embeddings for various downstream tasks.
In recommender systems, these techniques can be applied to a variety of graph structures, such as social networks~\cite{he2010social}, user-item bipartite graphs~\cite{he2016birank}, and knowledge graphs~\cite{wang2018ripplenet}, which are created using users, items, attributes, and other auxiliary information.
This section begins with an overview of graph representation learning techniques for embeddings, including spectral and spatial Graph Neural Networks (GNNs), graph pooling techniques, and self-supervised learning approaches for graph embedding.
These methods form the foundation for various graph-based embedding approaches in recommender systems.
Following this, we explore embedding methods based on various graph data structures: 1) Homogeneous Graphs; 2) Bipartite Graphs;
3) Heterogeneous Graphs; 4) Hypergraphs. Finally, we review relevant surveys and discuss future directions in this field. The techniques and models for each graph structure are summarized in Table~\ref{tab:graph_embedding_final}.%

\begin{table}[t!]
\centering
\caption{Graph-Based Embedding}
\label{tab:graph_embedding_final}
\begin{tabular}{>{\raggedright}m{2.5cm}|>{\raggedright}p{5.5cm}|m{6cm}}
\toprule
\multirow{12}{2.5cm}{\raisebox{-9\height}{\textbf{\begin{tabular}{@{}l@{}}Graph-Based \\ Embedding\end{tabular}}}} & 
\multirow{4}{*}{\raisebox{-3\height}{\textbf{\begin{tabular}{@{}l@{}}Graph Representation Learning \\ Techniques (\S\ref{ssec:graph_repr_learning})\end{tabular}}}} & 
\textbf{Spectral GNNs:} SpectralCF~\cite{zheng2018spectral}, JSCN~\cite{liu2019jscn}, SComGNN~\cite{luo2024spectral} \\
& & \textbf{Spatial GNNs:} Message-Passing Mechanism~\cite{kipf2016semi,gilmer2017neural}, Graph Transformers~\cite{dwivedi2020generalization, rampavsek2022recipe} \\
& & \textbf{Graph Pooling:} DiffPool~\cite{ying2018hierarchical}, SR-GNN~\cite{wu2019session}, RGNN~\cite{liu2021learning}, HG-Pool~\cite{wu2021user} \\
& & \textbf{Graph SSL:} SGL~\cite{wu2021SGL}, HHGR~\cite{zhang2021HHGR}, G-BERT~\cite{shang2019g-bert}, PT-GNN~\cite{hao2021PTGNN}, BUIR~\cite{lee2021bootstrapping} \\
\cmidrule(l){2-3}
& 
\multirow{2}{*}{\raisebox{-1\height}{\textbf{\begin{tabular}{@{}l@{}}Embedding of Homogeneous Graphs \\ (\S\ref{ssec:homogeneous})\end{tabular}}}} & 
\textbf{Random Walk Models:} Deepwalk~\cite{perozzi2014deepwalk}, APP~\cite{zhou2017scalable}, InfoWalk~\cite{zhou2021direction} \\
& & \textbf{GNN-based Models:} HyperSoRec~\cite{wang2021hypersorec}, M2GRL~\cite{wang2020m2grl}, DG-ENN~\cite{guo2021dual} \\
\cmidrule(l){2-3}
& 
\multirow{2}{*}{\raisebox{-2\height}{\textbf{Embedding of Bipartite Graphs (\S\ref{ssec:bipartite})}}} & 
\textbf{GCN-based Models:} GC-MC~\cite{berg2017graph}, STAR-GCN~\cite{zhang2019star}, NGCF~\cite{wang2019neural}, LightGCN~\cite{he2020lightgcn}, UltraGCN~\cite{mao2021ultragcn} \\
& & \textbf{Similarity-based Models:} Collaborative Similarity Embedding (CSE)~\cite{chen2019collaborative}, GE~\cite{xie2016learning} \\
\cmidrule(l){2-3}
& 
\multirow{2}{*}{\raisebox{-1\height}{\textbf{\begin{tabular}{@{}l@{}}Embedding of Heterogeneous Graphs \\ (\S\ref{ssec:heterogeneous})\end{tabular}}}} & 
\textbf{Social Recommendation:} DiffNet~\cite{wu2019neural}, GraphRec~\cite{fan2019graph}, DiffNet++~\cite{wu2020diffnet++}, DANSER~\cite{wu2019dual} \\
& & \textbf{Knowledge Graph-based:} KGCN~\cite{wang2019knowledge}, KGAT~\cite{wang2019kgat}, KGIN~\cite{wang2021learning} \\
\cmidrule(l){2-3}
& 
\textbf{Embedding of Hypergraphs (\S\ref{ssec:hypergraphs})} & 
\textbf{Key Models:} IHGNN~\cite{cheng2022ihgnn}, HyperGroup~\cite{guo2021hierarchical}, HEMR~\cite{la2022music} \\
\bottomrule
\end{tabular}
\end{table}
\begin{figure}[htbp]
\centering
\includegraphics[scale=0.125]{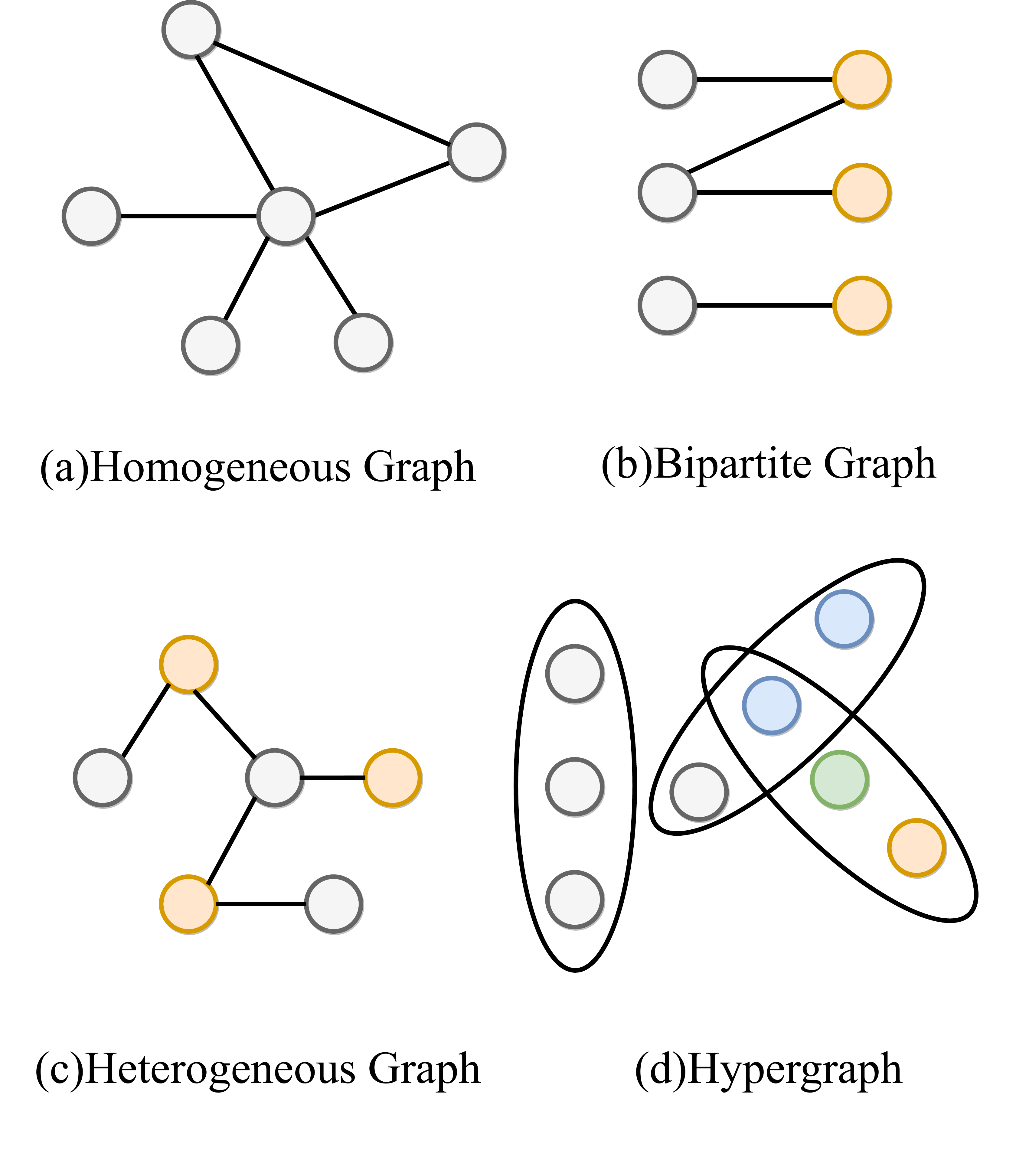}
\caption{Graph types in the recommendation system. (a) Homogeneous Graph: all nodes are of the same type. (b) Bipartite Graph: A unique structure with two node sets, where edges connect nodes from different sets. (c) Heterogeneous Graph: Encompasses diverse node categories that can be interconnected. (d) Hypergraph: In a hypergraph, each hyperedge can connect arbitrarily many nodes.}
\end{figure}
\subsection{Graph Representation Learning Techniques}
\label{ssec:graph_repr_learning}
\subsubsection{Spectral GNNs}

Spectral methods in graph theory are based on the eigendecomposition of graph matrices, particularly the graph Laplacian.
These methods provide a theoretical foundation for understanding graph structures and have been extended to develop Spectral Graph Neural Networks (Spectral GNNs)~\cite{bruna2013spectral,bo2023survey}.
Given an undirected graph $G$ with $n$ nodes, let $A$ be the adjacency matrix and $D$ be the diagonal degree matrix.
The unnormalized graph Laplacian is defined as $L = D - A$ and the normalized Laplacian is given by:  $L_{sym} = I - D^{-1/2}AD^{-1/2}$.
Spectral graph convolution is defined in the Fourier domain. Let $U$ be the matrix of eigenvectors of $L$, and $\Lambda$ be the diagonal matrix of its eigenvalues.
The graph Fourier transform of a signal $x$ is defined as $F(x) = U^Tx$, and the inverse transform as $F^{-1}(\hat{x}) = U\hat{x}$.
The spectral convolution of the signal $x$ with a filter $g_\theta$ is given by:

\begin{equation}
g_\theta * x = Ug_\theta(\Lambda)U^Tx
\end{equation}

where $g_\theta(\Lambda)$ is a diagonal matrix with entries that are a function of the eigenvalues.
In the context of recommender systems, this operation allows us to filter graph signals (e.g., user preferences) in the spectral domain, effectively capturing smooth variations across the user-item interaction graph.
Recent advancements in Spectral GNNs have primarily focused on enhancing computational efficiency and analyzing expressive power~\cite{bo2023survey}.
These theoretical developments have paved the way for practical applications in various domains, including recommender systems.
In this context, spectral methods have demonstrated their effectiveness across different recommendation scenarios.
SpectralCF~\cite{zheng2018spectral} applied spectral convolution to user-item bipartite graphs, improving cold-start recommendations.
Extending to cross-domain scenarios, JSCN~\cite{liu2019jscn} proposed transferable spectral representations and joint spectral convolution on multiple graphs, incorporating a domain adaptive module to handle domain incompatibility.
Further exploring the spectral properties of graphs, SComGNN~\cite{luo2024spectral} analyzed complementary item relationships in item graphs, identifying low-frequency and mid-frequency components in the spectral domain corresponding to relevance and dissimilarity attributes, respectively.
This observation informed the design of spectral filters for modeling these relationships.
Collectively, these studies illustrate the versatility of spectral methods in capturing complex interactions within recommendation tasks, highlighting their potential for future developments in the field of recommender systems.
\subsubsection{Spatial GNNs}
Spatial methods in Graph Neural Networks (GNNs) have gained significant popularity due to their intuitive design, scalability, and flexibility in handling diverse graph structures.
These methods typically employ a message-passing mechanism, iteratively updating node representations by aggregating information from neighboring nodes \citep{kipf2016semi,gilmer2017neural}.
The general update rule for node $i$ in the $k$-th layer can be expressed as:
\begin{equation}
\label{mpnn}
\mathbf{x}_i^{(k)} = \operatorname{Aggregation}^{(k)} \left( \mathbf{x}_i^{(k-1)}, \bigoplus_{j \in \mathcal{N}(i)} \operatorname{Message}^{(k)}\left(\mathbf{x}_i^{(k-1)}, \mathbf{x}_j^{(k-1)}, \mathbf{e}_{j,i}\right) \right),
\end{equation}
where $\mathbf{x}_i^{(k)}$ represents the updated node representation, $\mathcal{N}(i)$ denotes the neighborhood of node $i$, $\operatorname{Message}^{(k)}$ computes messages based on node and edge features, and $\operatorname{Aggregation}^{(k)}$ combines the current representation with aggregated messages.
The operator $\bigoplus$ represents the aggregation function. A complete GNN typically stacks multiple such layers, denoted as $\mathrm{GNN}(\cdot)$ for brevity.
Spatial methods capture and propagate local structural information across graphs, making them well-suited for recommender systems where user-item interactions are crucial.
These methods have been widely adopted in various recommendation scenarios, including collaborative filtering \cite{wang2019neural}, session-based recommendations \cite{wu2019session}, and social recommendations \cite{fan2019graph}.
Their flexibility allows the incorporation of heterogeneous data, enhancing the modeling of complex recommendation scenarios.
Recent advances in spatial GNNs span both theoretical and architectural fronts.
Theoretical work has focused on understanding GNN expressive power through the Weisfeiler-Leman (WL) hierarchy, proposing advanced methods to overcome $k$-WL limitations \cite{xu2018powerful,wang2023empirical,zhang2023complete,zhao2025graph}.
Architecturally, Graph Transformers have emerged as a powerful framework that combines message passing with global attention mechanisms \cite{dwivedi2020generalization, rampavsek2022recipe,kim2022pure,zhao2023hst}.
In recommender systems, transformer-based graph models have been developed to capture long-range dependencies in user-item graphs \cite{xia2022multi} and model complex multi-hop relationships \cite{li2023graph}, effectively addressing the over-smoothing problem of deep GNNs while maintaining scalability for large-scale recommendation scenarios.
\subsubsection{Graph Pooling}

In recommender systems, graph pooling is an important technique for obtaining global abstract representations of complex relational structures, enabling the capture of structural patterns in user preferences, item characteristics, and their diverse interactions.
This technique reduces graph complexity while preserving essential structural information, leading to more efficient and expressive recommendation models.
Graph pooling enhances the ability to model multi-scale relationships and aggregate information across different granularities, which is particularly valuable for handling the heterogeneous and multi-faceted nature of recommendation data.
Graph pooling typically aggregates node information to form more compact graph embeddings, emphasizing important structural patterns and relationships~\cite{wang2024graph}.
For example, DiffPool~\cite{ying2018hierarchical} learns soft assignments of nodes to clusters:
\begin{equation}
    \mathbf{S}^{(l)} = \text{softmax}(\text{GNN}_{l,\text{pool}}(\mathbf{A}^{(l)}, \mathbf{H}^{(l)}))
\end{equation}
where $\mathbf{S}^{(l)}$ is the assignment matrix, $\mathbf{A}^{(l)}$ is the adjacency matrix, and $\mathbf{H}^{(l)}$ is the node feature matrix at layer $l$.
These pooling operations enhance GNNs' ability to capture multi-scale structural information.
In the field of graph learning, the development of more powerful pooling techniques that incorporate insights from graph theory~\cite{bianchi2020spectral,tsitsulin2023graph} and topology~\cite{ying2024boosting}, while not tailor-made for recommender systems, still offers a potential path forward.
Graph pooling techniques have proven particularly valuable in recommender systems by enabling multi-scale relationship modeling.
For capturing short-term preferences, SR-GNN~\cite{wu2019session} employs pooling to aggregate item embeddings within sessions.
Extending beyond session-based scenarios, RGNN~\cite{liu2021learning} introduces a personalized graph pooling operator to learn hierarchical representations from review graphs, thereby constructing semantic representations for both users and items.
Most comprehensively, HG-Pool~\cite{wu2021user} represents each user as a personalized heterogeneous graph and proposes a novel heterogeneous graph pooling method, allowing the model to capture diverse preference patterns across different entity types and relations.
\subsubsection{Graph Embedding via Self-supervised learning}

\begin{figure}[htbp]
\centering
\includegraphics[scale=0.125]{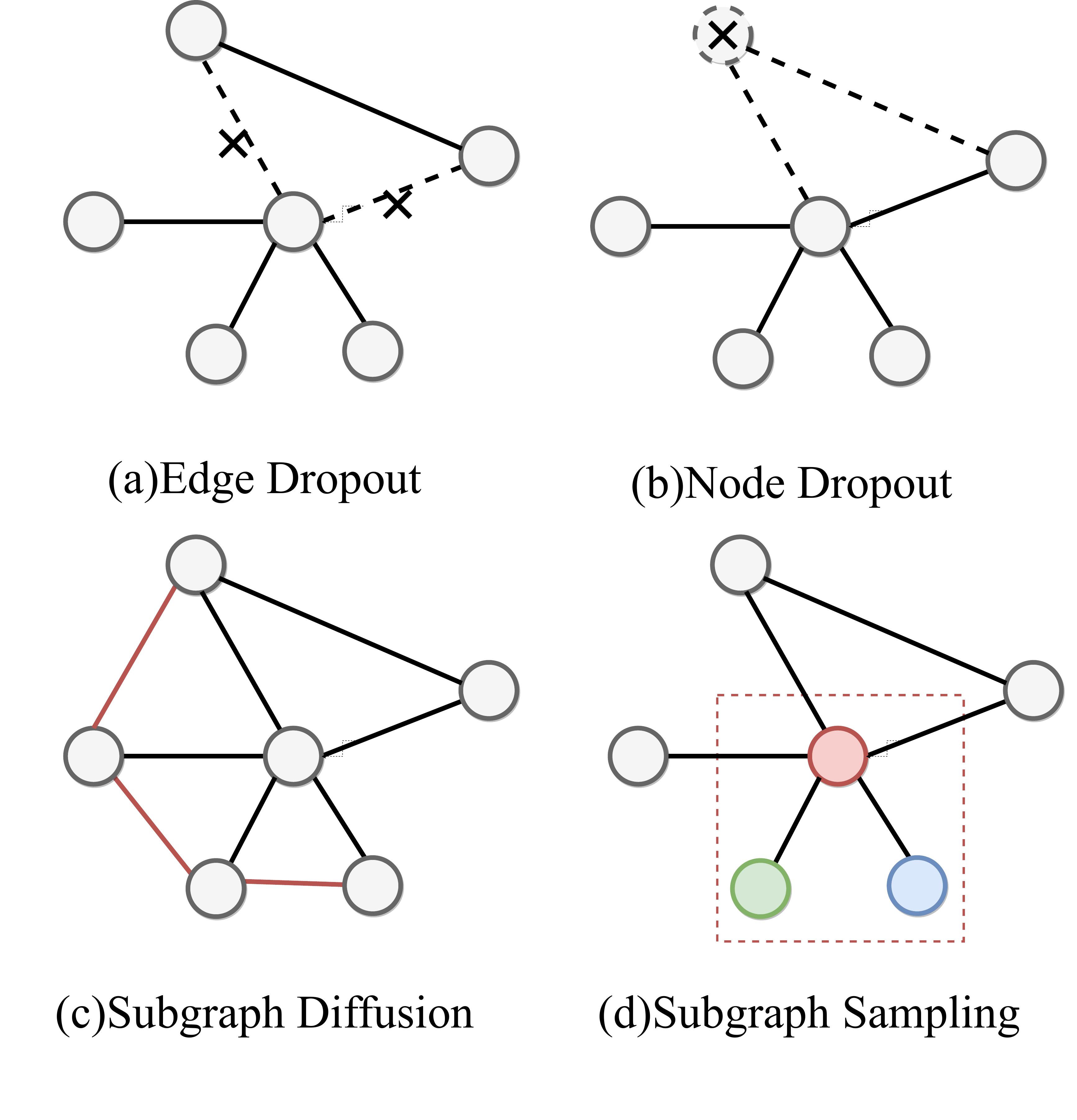}
\centering
\caption{Augmentation methods of Graph. Drop (a) interaction edges or (b) user/item nodes to identify influential components. (c) Introducing user-item similarity through new edges. (d) Sample local nodes and edges in the subgraph to accentuate connectivity.}
\label{seqGraph}
\end{figure}

Self-supervised learning (SSL) has emerged as a powerful paradigm for leveraging unlabeled data in machine learning models.
Building upon the foundations of spectral and spatial graph neural networks, Graph SSL (GSSL) techniques have shown promise in learning invariant and robust representations without extensive labeled data \citep{liu2022graph,xie2022self}.
Several prominent GSSL frameworks have been proposed in recent years, each with its unique approach.
GRACE \citep{zhu2020deep} uses edge perturbation and the InfoNCE estimator, while MVGRL \citep{hassani2020contrastive} introduces graph diffusion as augmentation and employs the Jensen-Shannon estimator as objective.
GCC \citep{qiu2020gcc} focuses on instance-level pretraining with subgraph sampling, and BGRL \citep{thakoor2021bootstrapped} adapts the BYOL approach to graphs without relying on negative samples.
G-BT \citep{bielak2022graph} applies the Barlow Twins objective to learn invariant graph representations.
These frameworks utilize augmentation techniques such as node/edge dropout, graph diffusion, and subgraph sampling to generate diverse graph views for SSL.
Some recent work has explored GSSL based on spectral principles, focusing on how edge perturbation can be guided by the spectrum~\cite{lin2023spectral,chen2024polygcl,yang2024spectral,jian2025rethinking}.
These generalized GSSL techniques show great promise in various graph-based tasks, and they open up new avenues for applications in recommender systems to exploit the rich structural information in various graphs.
In the following, we explore GSSLs adapted and extended for recommendation tasks, specifically categorized into contrastive, generative, and predictive approaches~\cite{yu2023self}.
Each category offers unique advantages in capturing different aspects of graph structures to enhance recommendation performance.
\paragraph{Contrastive Learning Methods}
Contrastive learning (CL) in recommendation systems aims to generate self-supervised signals by minimizing the distance between views of the same user or item and maximizing that between different users or items in the embedding space.
In graph-based recommendations, CL methods typically involve three key components: data augmentation, a graph encoder, and CL loss.
Common graph augmentation techniques in recommendation contexts, as illustrated in Figure \ref{seqGraph}, include:
\textbf{1) Node/Edge Dropout}: Removing user/item nodes or interaction edges to identify influential components and enhance system robustness.
\textbf{2) Graph Diffusion}: Introducing user-item similarity through new edges to model potential preferences.
\textbf{3) Subgraph Sampling}: Fabricating an augmented graph by selecting nodes and edges to accentuate local connectivity in the user-item graph.

The contrastive learning task can be formulated as:

\begin{equation}
f_{\theta}^* = \arg\min_{f_{\theta}, g_{\phi}} \mathcal{L}_{cl}\left(g_{\phi}\left(f_{\theta}(\mathcal{V}_1), f_{\theta}(\mathcal{V}_2)\right)\right)
\end{equation}
where $\mathcal{V}_1$ and $\mathcal{V}_2$ are two views of the graph, $f_{\theta}$ is a GNN encoder, $g_{\phi}$ is a projection head, and $\mathcal{L}_{cl}$ is a contrastive loss function.
Building upon traditional augmentation techniques, GraphAug \cite{zhang2024graph} proposes a more principled approach to graph augmentation through information bottleneck regularization.
Unlike conventional methods that rely on random perturbations, GraphAug automatically distills informative self-supervision information while filtering out noise, alleviating the fundamental challenge of noisy self-supervised signals in real-world recommendation environments.
Several notable works have applied contrastive learning to recommendation tasks.
SGL \cite{wu2021SGL} learns user and item embeddings in bipartite graphs using various augmentation techniques, while HHGR \cite{zhang2021HHGR} proposes a dual-scale node-dropping strategy specifically for group recommendations.
In the realm of cross-domain recommendations, CCDR \cite{xie2022CCDR} and PCRec \cite{wang2021PCRec} leverage contrastive learning to transfer knowledge between domains.
DCL \cite{liu2021DCL} challenges conventional assumptions about negative samples by conducting subgraph sampling through perturbation of a given node's L-hop ego network.
Heterogeneous graphs, which contain diverse node and edge types common in complex recommender systems, have also been explored in recent contrastive learning works \cite{wang2021self,zhu2022structure,cai2022heterogeneous,chen2023heterogeneous}.
These studies investigate how to leverage heterogeneous node and edge types to better reflect the multi-faceted nature of user preferences.
\paragraph{Generative Methods}
Inspired by the Masked Language Model (MLM) paradigm, notably successful in models like BERT \cite{devlin2018bert}, generative methods in this context often leverage the powerful ability of sequence models to learn embeddings by reconstructing corrupted data (e.g., sequences with masked items).
{BERT4Rec} \cite{sun2019bert4rec} utilizes the BERT \cite{devlin2018bert} encoder for sequential recommendation. In particular, it uses BERT to model sequential user behaviors.
The generative SSL task can be formulated as:

\begin{equation}
f_{\theta}^* = \arg\min_{f_{\theta}, g_{\phi}} \mathcal{L}_{gen}\left(g_{\phi}\left(f_{\theta}(\tilde{\mathcal{D}})\right), \mathcal{D}\right)
\end{equation}

where $\tilde{\mathcal{D}}$ is a corrupted version of the original data $\mathcal{D}$, and $\mathcal{L}_{gen}$ is typically a reconstruction loss such as cross-entropy or mean squared error.
Applying MLM-style generative self-supervised learning directly to graph-structured data can be less straightforward than on purely sequential data.
Consequently, many approaches in recommender systems first extract or transform graph data into sequences (e.g., user interaction sequences, node paths) and then apply sequence-based generative SSL techniques.
    For example, {G-BERT} \cite{shang2019g-bert} deals with the training data such that diagnosis and medical ontology are represented by a tree structure.
The model utilizes the graph attention network \cite{velivckovic2017graphattentionnetwork} to learn the ontology embedding from diagnosis and medical tree-like graphs in order to make medication recommendations.
    Then, it feeds the ontology embeddings into a BERT-like model and pre-trains it with MLM-based tasks, which include the self-prediction task and the dual-prediction task.
The former task is to predict the masked codes from the same type (i.e., medication or diagnosis) of ontology embeddings, and the latter is to reconstruct the masked codes from different types of ontology embeddings.
{
{PMGT}~\cite{liu2021PMGT} focuses on multimodal recommendations, handling nodes with diverse information in a multimodal graph (e.g., item descriptions and images).
They propose MCNSampling, a novel method to select neighboring nodes using importance scores and use BERT-like patterns for node embedding learning.
PT-GNN \cite{hao2021PTGNN}, instead of reconstructing masked nodes or items, addresses cold-start nodes with few neighbors using a meta-learning approach.
It first pre-trains a transformer encoder on the embeddings of the first $K$ neighboring nodes to simulate cold-start scenarios.
Then, a GCN model is pre-trained, incorporating the meta embedding in each convolution step.
It finally maximizes cosine similarity between the predicted embedding and the ground-truth embedding generated by NCF \cite{he2017neuralNCF} using all neighbor nodes.}

In conclusion, these generative methods leverage the successful MLM paradigm and the powerful modeling capabilities of Transformer encoders, aiming to learn high-performance and high-capacity embeddings.
While MLM is naturally suited to sequential data, researchers have often adapted these techniques for graph embedding learning by transforming graph data into sequential representations.
\paragraph{Predictive Methods}
Predictive SSL methods in recommendations generate new samples or labels from the original data to guide the learning process. These can be further divided into sample-based and pseudo-label-based approaches.

For pseudo-label-based methods, the task can be expressed as:

\begin{equation}
f_{\theta}^* = \arg\min_{f_{\theta}, g_{\phi}} \mathcal{L}_{pred}\left(g_{\phi}\left(f_{\theta}(\mathcal{D})\right), \hat{\mathcal{Y}}\right)
\end{equation}
where $\hat{\mathcal{Y}}$ represents generated pseudo-labels, and $\mathcal{L}_{pred}$ is a prediction loss function.

The core of Predictive Methods is the construction of pseudo-labels, and CHEST~\cite{wang2023curriculum} proposes the use of meta-path type as a prediction target in heterogeneous user-item graphs scenarios. BUIR~\cite{lee2021bootstrapping}, inspired by BYOL~\cite{grill2020bootstrap}, takes the representational consistency of the outputs of two encoders as a prediction task.

In conclusion, SSL techniques for graph embeddings extend the capabilities of spectral and spatial GNNs to scenarios with limited labeled data in recommender systems. By leveraging the unique graph structures inherent in recommender systems, these methods enhance the quality and robustness of recommendations. They capture complex relationships and implicit feedback, which are abundant in recommendation scenarios but difficult to model with traditional supervised approaches. 

\subsection{Embedding of Homogeneous Graphs}
\label{ssec:homogeneous}
Homogeneous graphs are the most basic graph structures, where all nodes in a homogeneous graph are of the same type.
In recommender systems, user social relationships or potential connections of items can be naturally modeled as homogeneous graphs, where nodes can be represented as users or items, and edges between nodes can represent the friendship between users or denote different items purchased by the same user.
Homogeneous graph-based graph learning algorithms (e.g., DeepWalk) can often be extended to more complex graph structures.
Deepwalk~\cite{perozzi2014deepwalk} is one of the most fundamental graph learning algorithms, which is based on the random walk algorithm to sample a sequence of nodes in the graph and encode them using skip-gram~\cite{mikolov2013efficient}.
Skip-gram is a classical word2vec model that maximizes the likelihood of central words to generate context words.
In Deepwalk, the central and context words represent a node $v_i$ and $k$ nodes that co-occur with node $v_i$, where $k$ is the window size of the skip-gram model.
The objective function of Deepwalk can be formulated as follows:

\begin{equation}
\max_{\theta} \quad \sum_{i=1}^N \sum_{-k \leq j \leq k, \neq 0} \log p\left(v_{i+j} \mid \mathbf{h}_{v_i}\right)
\end{equation}
Where $h_{v_{i}}$ denotes the embedding of node $v_i$, $\theta$ denotes the parameters of the neural network.
In this way, Deepwalk captures the structural information of the graph and generates similar embeddings for neighboring nodes.
Asymmetric Proximity Preserving (APP) graph embedding~\cite{zhou2017scalable} points out that in many downstream recommender system applications, the nodes in the graph do not have symmetry.
For example, the probability of observing a user purchasing a computer $i_{c}$ before buying a mouse $i_{m}$ is much higher than that of seeing a user buying a mouse $i_{m}$ before purchasing a computer $i_{c}$.
Deepwalk is based on skip-gram and needs to predict the context of each node, and cannot capture the asymmetry of the nodes. APP graph embedding only allows gradient updates in the direction of the sampled path.
InfoWalk~\cite{zhou2021direction} pointed out that there are nodes without in-degree in the directed graph, that is, dangling nodes, which cannot be handled by APP graph embedding because it cannot access such nodes.
However, as the graph size increases, the number of possible different node sequences generated by random walk-based methods becomes not handleable.
Some works try to solve this problem, such as LINE~\cite{tang2015line} and GNNs (e.g., GCN~\cite{kipf2016semi}, GAT~\cite{velivckovic2017graph}), which are widely used in recommender systems.
GNNs do not need to sample node sequences; they can be directly applied to the whole graph.
To learn high-quality graph embedding that can be applied to downstream tasks of recommender systems, GNNs need to be extended to take the special properties of recommender system tasks.
HyperSoRec~\cite{wang2021hypersorec} proposes that the user-user graph can be approximated as a structure of an $n$-order tree graph.
Chami et al.~\cite{chami2019hyperbolic} show that tree graph embedding in hyperbolic space can achieve better performance than its counterpart in Euclidean space, which may suffer from severe distortion.
HyperSoRec develops a hyperbolic mapping layer to map graph embedding in Euclidean space to hyperbolic space.
M2GRL~\cite{wang2020m2grl} proposes to combine homogeneous graphs of multiple views to enhance sparse features and enrich node information.
It constructs catalog view, instance view, and shop view, then adopts random walk and SGNS~\cite{levy2014neural} to learn an independent embedding of each view, and finally, maps them to a shared embedding space and aligns them separately.
DG-ENN~\cite{guo2021dual} also uses more than one view of homogeneous graphs to learn embeddings jointly.
It first learns embeddings from the user-user and item-item homogeneous graphs.
Then it leverages the user-item graph as a regularization to ensure the learned embeddings are smooth.
\subsection{Embedding of Bipartite Graphs}
\label{ssec:bipartite}
The bipartite graph is a unique structure with two node sets, where edges connect nodes from different sets.
This is commonly used in recommendation systems (e.g., user-item interaction graphs), represented as $\mathcal{G}=\left(\mathcal{V}_{A} \cup \mathcal{V}_{B}, \mathcal{E}\right)$, with $\mathcal{V}_{A}$ and $\mathcal{V}_{B}$ as node sets, and $\mathcal{E}$ denoting the edges.
Graph Convolutional Matrix Completion (GC-MC)~\cite{berg2017graph} predicts ratings using a graph encoder to aggregate information from neighboring nodes with uniform weight.
It aggregates user embeddings from one-hot embeddings based on interacted items, and item embeddings from one-hot vectors of interacting users.
A bilinear decoder then reconstructs user-item ratings.
However, GC-MC has drawbacks: 1) It employs one-hot encoding, hindering new node processing and becoming inefficient as the graph size grows.
2) GC-MC overlooks node features. STAR-GCN~\cite{zhang2019star} resolves these by end-to-end learning of low-dimensional node embeddings.
It simulates embedding new nodes with a masked vector, training the model to reconstruct embeddings.
NGCF~\cite{wang2019neural} addresses GC-MC's second drawback by incorporating node features in its embeddings.
LightGCN~\cite{he2020lightgcn} argues that, as the inputs of the user and item stem from ID embeddings lacking semantic information, there is no need for nonlinear transformations.
Therefore, LightGCN introduces a straightforward and efficient aggregation method:
\begin{equation}
\mathbf{h}_{\mathrm{u}}^{(\mathrm{k}+1)}=\sum_{\mathrm{i} \in \mathcal{N}_{\mathrm{u}}} \frac{1}{\sqrt{\left|\mathcal{N}_{\mathrm{u}}\right|} \sqrt{\left|\mathcal{N}_{\mathrm{i}}\right|}} \mathbf{h}_{\mathrm{i}}^{(\mathrm{k})}
\end{equation}
Where $\mathcal{N}_{u}$
and $\mathcal{N}_{i}$ denote the number of neighbor nodes of user $u$ and item $i$.
The $\mathbf{h}_{u}^{(l)}$ and $\mathbf{h}_{i}^{(l)}$ denote the node embeddings of user $u$ and item $i$ in layer $l$.
Although LightGCN greatly simplifies GCN, it still requires a long training time because multilayer message passing dominates the training of the model.
The relatively slow training limits the application of LightGCN in real recommendation scenarios.
UltraGCN~\cite{mao2021ultragcn} further removes the multi-layer messaging process by directly optimizing the cosine similarity of nodes and neighbors to capture the higher-order collaborative signals between users and items, and it uses a negative sampling strategy to avoid oversmoothing.
Recognizing the inherent sparsity within user-item interaction graphs, Collaborative Similarity Embedding (CSE)~\cite{chen2019collaborative} suggests enhancing embeddings by incorporating higher-order similarities among nodes of the same type.
CSE introduces two modules, DSEmbed and NSEmbed, to capture user-item similarity and inter-item/inter-user similarity within a user-item bipartite graph.
DSEmbed employs an optimization scheme based on rating data to model the proximity between users and items.
This is achieved by maximizing the log-likelihood function of positive and negative samples.
Conversely, NSEmbed employs random walks and CBOW~\cite{mikolov2013efficient} to predict central nodes, enabling the modeling of higher-order neighbor relationships among users or items.
Another approach, GE~\cite{xie2016learning}, proposes joint training to address the sparsity observed in user-POI bipartite graphs.
This is achieved by incorporating multiple graph perspectives, where supplementary bipartite graphs (e.g., POI-POI, POI-Region) augment the model's efficacy by providing enriched information.
\subsection{Embedding of Heterogeneous Graphs}
\label{ssec:heterogeneous}
The heterogeneous graph is a more realistic representation of many real-world graphs than simplified graphs like homogeneous and bipartite graphs.
These graphs encompass diverse node categories that can be interconnected.
In contrast to homogeneous graphs with just one node type, heterogeneous graphs are more intricate and contain more abundant information.
In the context of recommender systems, incorporating diverse relationships can lead to more accurate learned entity embeddings.
For instance, combining a user-item bipartite graph with a social network of users can enhance the precision of predicting user preferences.
Recent research focuses on enhancing user embeddings in recommender systems by incorporating social influence modeling.
DiffNet~\cite{wu2019neural} and GraphRec~\cite{fan2019graph} tackle social and user-item networks separately. 
DiffNet oversimplifies social influence modeling by assuming uniform influence among a user's neighbors.
In contrast, GraphRec employs GAT~\cite{velivckovic2017graph} to better model real-world social influence by weighing friends' influence based on the similarity of their initial embeddings.
DiffNet++ \cite{wu2020diffnet++} builds upon DiffNet by introducing a unified framework that considers both social networks and user-item bipartite graphs.
It addresses the limitation of DiffNet's inability to distinguish varying degrees of influence from different neighbors through a multi-level attention mechanism.
GES~\cite{wang2018billion} leverages related side information, like brand, to aggregate nodes' embeddings for obtaining item embeddings in heterogeneous graphs.
However, these methods primarily account for static social influence. DANSER~\cite{wu2019dual} introduces a dual GAT to capture both static and dynamic influence, acknowledging that a user's impact on friends can vary based on items, leading to more realistic embeddings.

The knowledge graph is also a type of heterogeneous graph that is widely used in recommender systems.
Nodes in the knowledge graph represent entities, edges represent the relationships between entities, and edge attributes describe the nature of these relationships.
By accurately describing real-world relationships, the knowledge graph provides rich semantic information that can improve interpretability and alleviate the sparsity problem in recommender systems.
To learn entity embeddings in the knowledge graph, several translation-based models such as TransE~\cite{bordes2013translating}, TransH~\cite{wang2014knowledge}, and TransR~\cite{lin2015learning} have been utilized.
However, these approaches fail to fully utilize the topological and semantic information in the knowledge graph, which is precisely what GNNs excel at.
To overcome this, methods like KGCN~\cite{wang2019knowledge} and KGAT~\cite{wang2019kgat} apply GCN and GAT, respectively, to produce higher-quality embeddings.
Nevertheless, these methods disregard users' diverse intents when modeling user-item relationships, thus constraining embedding learning quality.
Addressing this, KGIN~\cite{wang2021learning} introduces an intent layer between users and their interacted items for finer-grained relationship modeling.

\subsection{Embedding of Hypergraphs}
\label{ssec:hypergraphs}
The hypergraph is a graph structure in which an edge can connect arbitrarily many nodes, and edges in a hypergraph are called hyperedges.
In recommender systems, hypergraphs can be used to model complex entity relationships.
IHGNN~\cite{cheng2022ihgnn} enhances node embeddings through hypergraphs, revealing higher-order interaction patterns in user query histories.
It introduces hyperedges $(u,q,p)$ for user query and product, employing three high-order feature interactions for hyperedge embedding aggregation.
HyperGroup~\cite{guo2021hierarchical} targets group recommendation, addressing potential misalignment between user and group preferences by representing groups as hyperedges.
It aggregates neighboring hyperedges to capture group similarity.
HEMR~\cite{la2022music} focuses on music recommendation using hypergraph embeddings.
It employs hyperedge-level random walks, followed by skip-gram for node embedding learning.
\subsection{Surveys and Future Directions}
\noindent \textbf{Surveys.} Unlike our discussion focusing on learning graph embedding in recommender systems, Wang et al.~\cite{wang2020graph} focus on the design of graph neural networks in different types of graph structures, while they do not consider the application of hypergraphs in recommender systems.
Gao et al.~\cite{gao2021graph} focus on the application of graph neural networks in different recommender scenarios such as sequence recommendation and multi-behavior recommendation.
Embeddings from static graphs can result in uninteresting recommendations for users.
While PGE~\cite{li2017learning} considers temporal decay for item weights, it still treats graphs as static and overlooks evolving network structures.
Leveraging dynamic graph neural networks~\cite{manessi2020dynamic,you2022roland} for effective recommender system applications is a critical area for future exploration.

\noindent \textbf{Future Direction 1: Fair Graph Embedding for Recommendation.}
The pursuit of fairness in machine learning is gaining considerable traction, with a critical goal of mitigating biases inherent in data-driven models.
In the context of graph-based recommender systems, preventing the learning and propagation of biased patterns is especially crucial due to their significant societal impact.
Thus, ensuring the fairness of learned embeddings is a pressing concern.
While initial studies in fair graph learning (e.g., GEAR~\cite{ma2022learning} for fair graph classification and FairGo~\cite{wu2021learning} for fair influence maximization) provide relevant foundations, the specific area of fair graph embedding tailored for recommendation systems remains relatively underexplored.
Developing robust methods to achieve fairness in these embeddings and establishing comprehensive benchmarks to evaluate fairness in graph-based recommendation systems are key research avenues.
For a broader understanding of fair machine learning, readers are referred to the survey by Mehrabi et al.~\cite{mehrabi2021survey}.

\noindent \textbf{Future Direction 2: Leveraging Topological Insights in GNNs for Enhanced Recommender Systems.}
Recent breakthroughs in topology-inspired graph neural networks~\cite{papillon2023architectures} present exciting opportunities for advancing recommender systems.
For instance, curvature-based graph rewriting techniques have demonstrated success in mitigating common GNN pitfalls like over-smoothing and over-squashing~\cite{topping2021understanding,nguyen2023revisiting,fesser2024mitigating}.
Applying these techniques could potentially lead to more nuanced and informative representations of user-item interactions, especially in complex recommendation scenarios.
Persistent Homology (PH)~\cite{edelsbrunner2008persistent}, a powerful tool from topological data analysis for capturing multi-scale topological features, has been effectively integrated with GNNs.
This integration has shown promising results and, in some cases, significant improvements in diverse graph-based tasks, including node classification~\cite{horn2021topological}, link prediction~\cite{yan2021link}, and the regression of molecular properties~\cite{immonen2024going}.
In recommender systems, applying PH to user-item interaction graphs could potentially uncover topological features (e.g., persistent cycles or connected components) that remain stable across various similarity thresholds.
These features might correspond to robust patterns in user preferences or item relationships, offering novel insights to enhance collaborative filtering algorithms.
Furthermore, the persistence diagrams generated through PH analysis could serve as unique topological signatures for users or items, thereby potentially enriching their feature representations in recommendation models.
Additionally, advanced structural coding schemes~\cite{zhao2021stars,zhang2023rethinking}, effective in identifying critical components within graphs, could be adapted to pinpoint influential users or pivotal items in recommendation networks.
Such an adaptation could improve the precision of social and collaborative filtering recommendations.

\section{Hash Embedding}
\label{sec:hash_embedding}


{
In recommender systems, an alternative to the prevalent one-hot encoding for representing categorical attributes is hashing. Hashing embeddings were introduced to address the limitations associated with one-hot encoding. One-hot encoding often results in high-dimensional and extremely sparse feature matrices, posing significant scalability challenges, especially within complex deep recommender systems. Hashing embeddings provides an effective solution to these challenges, offering benefits for both scalability enhancement and complexity reduction.
Hashing embedding involves the application of one or more hash functions to the original sparse encoding, effectively curbing storage demands and computational overhead. Compared to raw one-hot encodings, hash embeddings manage to maintain recommendation effectiveness while substantially reducing the complexity of training and inference.}

\begin{table}[t!]
\centering
\caption{Hash Embedding}
\label{tab:hash_embedding_final}
\begin{tabular}{>{\raggedright}m{2.5cm}|>{\raggedright}p{5.5cm}|m{6cm}}
\toprule
\multirow{3}{2.5cm}{\raisebox{-6\height}{\textbf{Hash Embedding}}} & 
\textbf{Single Function Hash Embedding (\S\ref{ssec:single_hash})} & 
\textbf{The Hashing Trick:} (Yahoo)~\cite{weinberger2009feature} \\
\cmidrule(l){2-3}
& 
\textbf{Multiple Functions Hash Embedding (\S\ref{ssec:multi_hash})} & 
\textbf{Key Models:} Bloom Filter based~\cite{bonomi2006improved}, Hash Embedding~\cite{tito2017hash}, Hybrid Hashing (Twitter)~\cite{zhang2020model}, Q-R Trick (Facebook)~\cite{shi2020compositional} \\
\cmidrule(l){2-3}
& 
\textbf{DHE: Dense Hash Embeddings (\S\ref{ssec:dhe_hash})} & 
\textbf{Key Models:} Dense Hash Embedding (DHE)~\cite{kang2020learning}, Dynamic Sparse Learning (DSL)~\cite{wang2024dynamic} \\
\bottomrule
\end{tabular}
\end{table}


Considering various hash embedding methods based on factors like the number of hash functions, their utilization, and post-processing steps, we categorize the literature into three groups: 1) single function hash embedding; 2) multiple functions hash embedding; 3) dense hash embedding. A comparison of these categories is presented in Table~\ref{tab:hash_embedding_final}.
\subsection{Single Function Hash Embedding}
\label{ssec:single_hash}
The hashing trick, proposed by Yahoo~\cite{weinberger2009feature}, uses a single hash function to map features to hash values.
This method is particularly relevant in collaborative filtering scenarios, exemplified by matrix factorization ($\mathbf{M} = \mathbf{U}\mathbf{V}^T \in \mathbb{R}^{m\times n}$), where $\mathbf{U} \in \mathbb{R}^{m\times d}$ and $\mathbf{V}$ $\in \mathbb{R}^{n\times d}$ are factorized matrices of the sparse $\mathbf{M}$. The hashing trick diminishes the dimensions of these matrices through a binary encoding function $f_E$ and a hash function $f_H$. 
In this context, with feature index $(j,k)$, and position index $p$, $f_E$ transforms $(j,k)$ into values in $\{0,1\}$, effectively positioning the feature in hash table position $i$ using hash function $f_H$. This leads to $\mathbf{U'}$ $\in$ $\mathbb{R}^{{m'}\times d}$ and $\mathbf{V'}$ $\in$ $\mathbb{R}^{n' \times d}$, where $m' \ll m $ and $ n' \ll n $, as follows:

\begin{equation}
\left \{
\begin{aligned}
& \mathbf{U'}_{pk} = \sum_{j:f_H(j,k)=p}f_E(j,k)\mathbf{U}_{jk} \\
& \mathbf{V'}_{pk} = \sum_{j:f_H'(j,k)=p}f_E'(j,k)\mathbf{V}_{jk}
\end{aligned}
\right.
\end{equation}
Here, $f_E'$ and $f_H'$ are separate functions. The hashing function maps feature values to $\{0,1,2,...,m'-1\}, m' \ll m$, while $f_E(j,k)=0$ if $f_H(j,k)\not = p$, and $1$ otherwise. 
It is equivalent to its matrix representation. For instance, consider the compression of $\mathbf{U}\in\mathbb{R}^{m \times d}$ into $\mathbf{W} \in \mathbb{R}^{m' \times d}$, where $m' \ll m$. This reduction simplifies the computation of the final hash embedding, as exemplified by the user vector $\mathbf{U}_{p,\cdot}$:
\begin{equation}
\begin{aligned}
\mathbf{U}_{p, \cdot}^T = \mathbf{W}^T \mathbf{H} \mathbf{e}_p
\end{aligned}
\end{equation}
In this equation, $\mathbf{H} \in \{0,1\}^{m' \times m}$ is a hash matrix, and $\mathbf{e}_p$ is a $d$-dimensional one-hot vector with its $p$-th element equals to $1$.
$h_{ij}$ of the hash matrix $\mathbf{H} \in \{0,1\}^{m' \times m}$ is defined as:
\begin{equation}
h_{ij} = \left \{
\begin{aligned}
& 1, \ \rm{if} \; j \ \rm{mod} \ m=i \\
& 0, \ \rm{otherwise} \\
\end{aligned}
\right.
\end{equation}
The hashing trick has had a profound impact on the field of hash embedding for recommendation systems. This approach stands as the pioneer of simple yet powerful techniques that markedly enhance the efficiency of original one-hot embeddings, particularly when dealing with high-dimensional features.

\begin{figure}[htbp]
    \centering
    \includegraphics[width=0.8\textwidth]{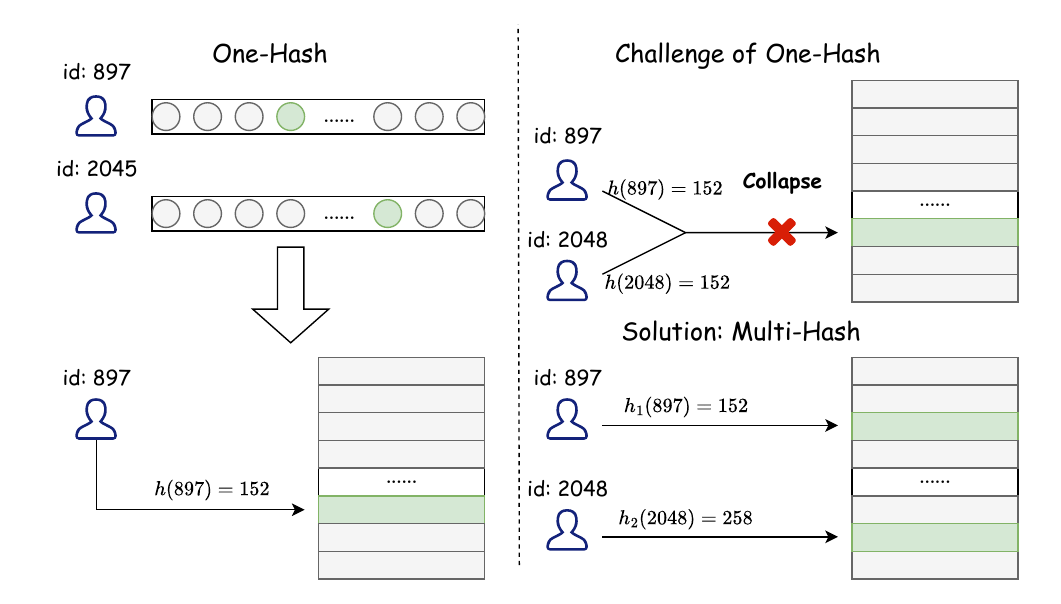}
    \caption{The mechanism and challenge of hash embeddings. (Left) The 
    "hashing trick" maps high-cardinality IDs to a smaller embedding table. 
    (Top-Right) A key challenge is hash collision, where different IDs 
    (e.g., 897 and 2045) collapse to the same index, causing representation ambiguity. 
    (Bottom-Right) Multi-hash methods provide a solution by using multiple 
    hash functions to ensure unique representations.}
    \label{fig:hash_collision_solution}
\end{figure}

The primary challenge of this "hashing trick" is the potential for hash 
collisions, where different features are mapped to the same embedding index, leading to information loss. To counteract this, multi-hash approaches have been proposed, as illustrated in Fig.~\ref{fig:hash_collision_solution}.

\subsection{Multiple Functions Hash Embedding}
\label{ssec:multi_hash}
To counteract the information loss stemming from hash collisions, approaches utilizing multiple hash functions have been devised. These methods can be categorized based on their encoding functions. The concept behind Embeddings with Multiple Hash Functions aligns with the hashing trick's modulus hashing and hash embedding table. 

A notable technique is the Bloom filter \cite{bonomi2006improved}, which employs multiple hash functions to map features into an embedding space. This process generates a binary embedding through multiple hash values, enabling intermediate products to be recovered or mapped back from the embedding to the final product without compromising information integrity.
Another innovation, hash embedding \cite{tito2017hash}, merges word embedding and the hashing trick. It entails computing the hash embedding as the product of embedding vectors and the corresponding element in the weight vector for each token. Unlike the hashing trick, which employs a solitary hash function, hash embedding deploys multiple functions. It initially queries the relevant embedding based on the feature index and subsequently employs a weighted sum of these embeddings.

Hybrid hashing \cite{zhang2020model}, introduced by Twitter, follows a similar premise. It employs two hash functions but applies them exclusively to infrequent feature values. For frequent feature values, the one-hot full embedding method is employed. This approach divides the embedding space considering feature frequencies, ensuring that vital information from the top-$K$ frequent features is retained. By employing a combination of one-hot full embedding and hybrid hashing, Facebook's Q-R trick \cite{shi2020compositional} adopts the complementary partition idea with multiple hash functions to avert hash collisions.
Q-R trick introduces complementary hash functions through two embedding tables: $\mathbf{W}_1 \in \mathbb{R}^{l \times d}$ and $\mathbf{W}_2 \in \mathbb{R}^{\frac{a}{l} \times d}$ for each categorical feature with $a$ distinct values. The final embedding results from pooling the outputs of each hash function. The hash embedding combines two components: $\mathbf{W}_1$ represents the coarse-grained aspect, where multiple values can map to the same row, while $\mathbf{W}_2$ functions as the fine-grained counterpart, compensating for collisions in $\mathbf{W}_1$. The ultimate hash embedding is computed as:
\begin{equation}
\begin{aligned}
\mathbf{U}_{p, \cdot}^T = \mathbf{W}_1^T \mathbf{H} \mathbf{e}_p \odot \mathbf{W}_2^T \mathbf{H}' \mathbf{e}_p
\end{aligned}
\end{equation}
Here, $\mathbf{H}$ and $\mathbf{H}'$ are the corresponding hash matrices, and $\odot$ is element-wise product. This design employs complementary hash functions to ensure representation uniqueness.

Recent comprehensive benchmarking studies have provided new insights into the effectiveness of hash embedding methods in real-world deployment scenarios. Tran et al.~\cite{tran2025thorough} conducted an extensive evaluation of lightweight embedding-based recommender systems, revealing that the performance of hash embedding methods varies significantly across different recommendation tasks. Their analysis shows that while hash embedding approaches like the Q-R trick achieve substantial memory reduction, the trade-offs between compression ratio and recommendation quality are highly task-dependent. The benchmark results indicate that simple magnitude-based pruning can sometimes outperform complex hash embedding schemes, suggesting that the field may benefit from focusing on fundamental compression principles rather than increasingly sophisticated hashing strategies.

\subsection{DHE: Dense Hash Embeddings}
\label{ssec:dhe_hash}

In contrast to the aforementioned methods, Google's Dense Hash Embedding (DHE) \cite{kang2020learning} introduces several significant innovations. This approach focuses on enhancing hash functions to generate denser embedding vectors. Notably, DHE deploys a DNN-based decoding layer in lieu of conventional embedding tables, leading to markedly superior performance compared to other hash embedding techniques. Remarkably, its performance can even rival that of one-hot encoding. However, this enhancement comes at the trade-off of a larger model size relative to other hash embedding methods.

DHE employs a multitude of hash functions, approximately $1,000$ in number, along with a DNN. These hash functions map categorical features like IDs to high-dimensional vectors (e.g., $1,024$ dimensions), with vector elements in the range of $\{1,2,...,m\}$. These vectors can be left normalized, drawn from a uniform distribution, or subjected to transformation into a normal distribution via the Box-Muller method \cite{box1958note}. Moreover, a DNN with $h$ layers and dropout regularization serves as a decoder. It maps the identity-encoded vectors of dimension $k$ to final hash embedding vectors of dimension $d$.
DHE stands out in its ability to maintain the uniqueness of representations compared to other original hash embedding methods. Furthermore, it addresses the out-of-vocabulary challenge by ensuring the feature embedding is influenced by changes in embedding net parameters. DHE also offers a mechanism for enhancing generalization. This involves concatenating different series of categorical features to the ID-type feature during the decoding process, further improving generalization capabilities.

Building on the dense hash embedding concept, recent work has explored dynamic approaches to hash embedding that can adapt to changing data distributions. Dynamic Sparse Learning (DSL) methods dynamically adjust the sparsity distribution of hash embeddings through pruning and growth strategies, allowing the hash structure to evolve during training~\cite{wang2024dynamic}. This adaptive approach addresses a key limitation of traditional hash embedding methods: their static nature, which cannot respond to shifts in user behavior patterns or item popularity distributions.

\subsection{Surveys and Future Directions}
\par \noindent \textbf{Survey.}
Kang et al. \cite{kang2020learning} mention many hashing embedding methods with various techniques in different application scenarios.
Ghaemmaghami et al. \cite{ghaemmaghami2022learning} present an overview of popular hash embedding models in deep recommendation systems. It primarily discusses the initial purpose of hash functions: compressing models while controlling hash collisions.
\par


{
\noindent \textbf{Future Direction: Enhancing Hash Embedding with Multiple Functions.}
In stark contrast to employing a solitary hash function, leveraging multiple hash functions or embracing binary encoding has proven to exert superior control over the scale of the embedding matrix, concurrently reducing the loss of information. Recent innovative designs highlight~\cite{kang2020learning} the beneficial impact of incorporating an augmented number of hash functions. Such a framework possesses the potential to seamlessly replace the extensive and costly hash embedding tables utilized in preceding methodologies, presenting a notably more efficient storage alternative.}

\section{AutoML \& Embedding}
\label{sec:automl_embedding}

    Auto machine learning (AutoML) is a process that automatically generates optimal machine learning solutions for repetitive and time-consuming tasks. In the embedding learning scenario, it could search for the best size of the embedding layer. 
    Concretely, the traditional deep learning approach employs a fixed, uniform embedding size for all features, disregarding the varying importance of each feature in recommendation tasks, which inevitably leads to suboptimal performance. {Additionally, the use of excessively large embedding sizes contributes to inflated storage usage and elevated computational expenses.  To counter these issues, the AutoML framework introduces the concept of tailoring appropriate embedding sizes for each feature. Neural Architecture Search (NAS) plays a pivotal role in addressing this concern within the realm of AutoML. NAS's substantial advancements now enable efficient exploration of optimal configurations, including embedding sizes, for recommendation systems within a feasible timeframe.}

    The methods discussed in this section will be classified by the kinds of search strategies in NAS, which include reinforcement learning \cite{kaelbling1996reinforcement}, gradient-based algorithm \cite{ruder2016gradient}, and evolutionary algorithm \cite{qin2008evolutionary}. In addition, considering that there are a few papers on the evolutionary algorithm, we introduce this category with other unique methods together. These search strategies and their corresponding models are detailed in Table~\ref{tab:automl_embedding_final}.

\begin{table}[t!]
\centering
\caption{\begin{tabular}{@{}c@{}}AutoML \& Embedding\end{tabular}}
\label{tab:automl_embedding_final}
\begin{tabular}{>{\raggedright}m{2.5cm}|>{\raggedright}p{5.5cm}|m{6cm}}
\toprule
\multirow{7}{2.5cm}{\raisebox{-4.5\height}{\textbf{\begin{tabular}{@{}l@{}}AutoML \& \\ Embedding\end{tabular}}}} & 
\textbf{Reinforcement learning based method (\S\ref{ssec:rl_automl})} & 
\textbf{Key Models:} NIS~\cite{joglekar2020nis}, ESAPN~\cite{liu2020esapn} \\
\cmidrule(l){2-3}
& 
\multirow{3}{*}{\raisebox{-2\height}{\textbf{Gradient-based Method (\S\ref{ssec:gradient_automl})}}} & 
\textbf{Soft Selection Models:} DNIS~\cite{cheng2020dnis}, autoSrh~\cite{wei2021autoias}, AutoEMB~\cite{zhaok2021autoemb}, AutoDim~\cite{zhao2021autodim} \\
& & \textbf{Pruning-based Models:} AMTL~\cite{yan2021amtl}, SSEDS~\cite{qu2022SSEDS} \\
& & \textbf{Budget-constrained Models:} BET~\cite{qu2024budgeted} \\
\cmidrule(l){2-3}
& 
\multirow{3}{*}{\raisebox{1\height}{\textbf{Other methods (\S\ref{ssec:other_automl})}}} & 
\textbf{Evolutionary Algorithms:} RULE~\cite{chen2021rule} \\
& & \textbf{Regularization/Pruning:} PEP~\cite{liu2021pep} \\
& & \textbf{Anchor Embeddings:} ANT~\cite{liang2020ant}, autoDis~\cite{guo2021autodis} \\
\bottomrule
\end{tabular}
\end{table}
\subsection{Reinforcement learning based method}
\label{ssec:rl_automl}

\begin{figure}[htbp]
    \centering
    \includegraphics[width=0.85\textwidth]{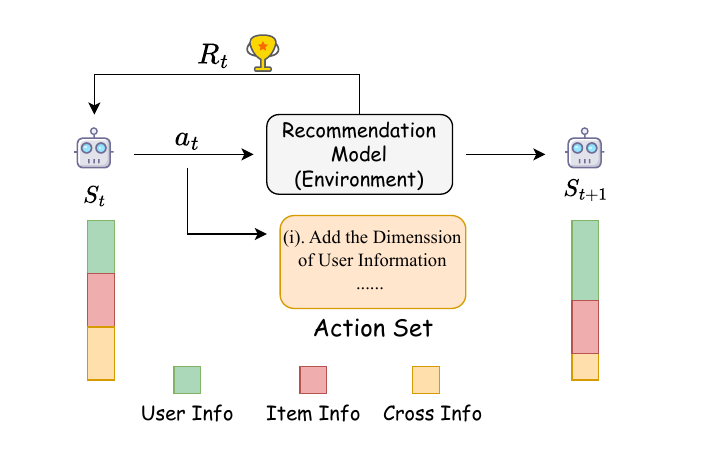}
    \caption{The reinforcement learning loop for Embedding Dimension Search (EDS). 
    The Controller (Agent), based on the current state \textit{S\textsubscript{t}}, 
    selects an action \textit{a\textsubscript{t}} (e.g., modifying embedding dimensions). 
    The Recommendation Model (Environment) executes this action and returns a 
    reward \textit{R\textsubscript{t}} and the next state 
    \textit{S\textsubscript{t+1}}. The agent learns by optimizing its policy to 
    maximize future rewards.}
    \label{fig:automl_rl_process}
\end{figure}

    Reinforcement learning is quite simple to grasp. Think of it like improving your game strategy: if a move or tactic leads to a higher score, reinforcing it enhances your performance. This learning approach involves five main parts: the environment (like a game scenario), the controller or player, actions, rewards, and status. In the case of embedding dimension search (EDS), consider a recommendation model as the setting, evaluating performance to determine rewards. The player, often represented by a policy network, decides what move to make. Rewards are based on how well the model performs. The player uses these rewards to decide how to adjust the current embedding size. The status refers to the player's situation, like the network parameters. The entire process, often referred to as Embedding Dimension Search (EDS), is illustrated in Fig.~\ref{fig:automl_rl_process}.

{{NIS} \cite{joglekar2020nis} initiates Embedding Dimension Search (EDS) by organizing features based on frequency, creating an embedding $\mathbf{E}$ that assigns larger sizes to frequent features. It divides $\mathbf{E}$ into matrices using a fixed search grid.
    Within this reinforcement learning framework, a recommendation model computes rewards, and a neural network-based player selects actions through softmax probabilities. The reward integrates memory costs and target objectives into a composite loss function. Asynchronous Advantage Actor-critic (A3C) \cite{A3Cmnih2016asynchronous} trains the player and recommendation model alternately.
{ESAPN}~\cite{liu2020esapn} dynamically searches for suitable embedding sizes for users and items in stream recommendation. At first, It establishes potential embedding size ranges. Using a recommendation model tailored for stream recommendations, it employs a reinforcement learning approach. The controller, comprising user and item multilayer perceptrons, uses frequency and current embedding size as inputs for the policy network. The controller's action decides whether to increase or maintain embedding sizes. The reward reflects its predictive capability and guides its decision-making process, which can be calculated from losses of the stream recommendation model as follows:}
    \begin{equation}
        R^{u/i}=\frac{1}{T}\sum_{t=1}^{T}L^{(u/i)}-L 
    \end{equation}
    where the input is a loss sequence of users or items with length $T$, $L^{(u/i)}_t$ represents the $t$-th prediction loss of user $u$ or item $i$, and $L$ is the loss of the current item or user. The specific loss function is suitable for the typical recommendation task.
    To optimize the parameters of the policy network according to the reward, the Bellman equation is formulated as follows:
    \begin{equation}
        J(\Phi)=\mathbb{E}_{a \sim \Phi(a \mid s)} R(a \mid s)
    \end{equation}
    where $\Phi$ is the parameter of the policy network for the user or item, $a$, and $s$ denote the action and state, respectively. However, it is hard to calculate its gradient in practice; thus, it uses Monte-Carlo sampling to evaluate the $\nabla_{\Phi} J(\Phi)$. 
    In the end, {inspired by ENAS}, it uses sampled validation data to optimize the policy network and train the recommendation model based on training data.
    
    In summary, reinforcement algorithm-based methods solve the non-differentiable problem of the hard selection in a discrete search space and are an efficient and effective search strategy to significantly reduce the search space in embedding size. However, the controller takes a hard selection in a pre-defined embedding candidate size, which is likely to achieve a sub-optimal performance. To get higher performance in embedding size selection, the soft selection strategy is widely adopted in gradient-based methods.

\subsection{Gradient-based Method}
\label{ssec:gradient_automl}

    Gradient-based methods, such as the Differentiable Neural Architecture Search (DARTS) method proposed by Liu et al. \cite{liu2018darts}, have gained attention. These methods transform the search space from discrete to continuous, and the optimization process is guided by gradient calculations or inspired by DARTS.
    {For example, DNIS \cite{cheng2020dnis} argues that a pre-defined discrete search space could limit the flexibility of the searching model. Thus, it optimizes the NIS approach by leveraging a soft selection layer to relax the search space to continuous space for performance improvement.  
    }

    Specifically, it denotes an embedding $E$ with max dimension K by a multi-hot embedding index vectors set $\mathcal{D}=\{d_1,\cdot\cdot\cdot,d_N\}$ and a values vectors set $\mathcal{V}=\{v_1,\cdot\cdot\cdot,v_K\}$ with the search space of $2^{NK}$. As same as the NIS, it divides the features into $S$ blocks based on the descending order of frequency. Thus, the search space is reduced from $2^{NK}$ to $2^{SK}$.
    After that, it defines the soft selection layer as a numerical matrix $\alpha \in \mathbb{R}^{S \times K}$ with elements in the range $[0,1]$, which aims to shrink the elements in embedding within a small number. 
    The purpose of this layer is to reduce the values within the embedding to a smaller range.
    The soft selection operation can be calculated as follows:
    \begin{equation}
        \widetilde{e}_i = e_i  { \odot}  \alpha_i
    \end{equation}
    where $\widetilde{e}_i$ is the $i$-th row of output embedding and $\alpha_i$ is $i$-th row of $\mathbf{A}$ matrix. In addition, matrix $\mathbf{A}$ is gained through learning. By inserting a soft selection layer between the embedding layer and interaction layer, it could be optimized jointly like the DARTS method. In particular, it defines a bi-level optimization problem that has been solved by DARTS. 

    Similar to DNIS, autoSrh \cite{wei2021autoias} follows the pipeline of DNIS to make tabular Data prediction. The difference is that it debates the idea that setting a pruning threshold could satisfy the requirement of limited memory usage.
    
    Moreover, to improve the hard selection in ESAPN, AutoEMB ~\cite{zhaok2021autoemb} proposed the soft selection layer as a weighted sum operation. 
    Concretely, similar to ESAPN, it defines some candidate embedding sizes for users and items, and 
    employs matrices to transform user and item embedding of candidate sizes into a uniform embedding size
    for plugging into deep learning-based recommendation models. However, the author noted that the{simple linear transformation} process results in significant variation in the values of embeddings with different sizes, rendering them incomparable.
    To address this issue, the approach applies Batch-Norm with Tanh activation to normalize the transformed user and item embedding vectors. Subsequently, it introduces two multiple perceptron networks to select the embedding size. For an end-to-end differentiable framework, it takes a selection layer to get the embedding size.

    However, there is another approach called AutoDim \cite{zhao2021autodim} that addresses the issue of non-differentiability by employing Gumbel-softmax tricks as a soft selection layer. AutoDim builds upon AutoEMB by extending its input from just users and items to various feature fields such as gender, age, and more. This expansion allows for more comprehensive embeddings that can capture additional contextual information for improved performance. The evolution of gradient-based embedding search has led to more sophisticated approaches that address real-world deployment constraints. Budgeted Embedding Table (BET)~\cite{qu2024budgeted}, which advances beyond traditional AutoML approaches by incorporating explicit memory budget constraints into the search process. Unlike previous methods that optimize embedding dimensions independently, BET formulates the problem as finding table-level actions that are guaranteed to meet pre-specified memory budgets while maximizing recommendation performance.

    Another gradient-based idea to find the optimal embedding size is to prune the original embedding through a mask matrix, which is learned from the training data. For example, AMTL \cite{yan2021amtl} utilizes an Adaptively-Masked Twins-based layer (AMTL) to generate the mask vector. Specifically, the input of AMTL is the frequency attribute of features, such that the rank of frequency in a given feature field and AMTL consists of two multiple perceptrons (i.e., h-AML and l-AML) for high-frequency and low-frequency data, respectively, to avoid the parameter update dominated by high-frequency samples. Then, the final output is the weighted sum to fuse the output of two branch AMLs.
    Lastly, the approach applies a differentiable temperature softmax function to the output, generating a probability mask vector for selecting an embedding size for high or low-frequency features. However, considering the high training costs of AMTL, SSEDS \cite{qu2022SSEDS} introduces a saliency criterion to determine the significance of each element in the embedding matrix. The saliency score is obtained through a single forward-backward propagation. Using the saliency score, the mask matrix is generated by retaining the top-K important parameters, taking into account memory constraints.

\subsection{Other methods}
\label{ssec:other_automl}

Evolutionary algorithms, inspired by natural evolution, have been used for optimizing neural network parameters since the twentieth century. In Neural Architecture Search (NAS), these algorithms involve creating an initial set of architectural designs (seed), combining and mutating them, and evaluating the performance of resulting designs. For instance, RULE~\cite{chen2021rule} employs an evolutionary algorithm to learn adaptable embeddings with distinct sizes for different items, akin to NIS.
In RULE, an estimator decides which designs to eliminate based on mixed embedding block data. The estimator is trained using these compositions and evaluated using metrics like recall.
Evolution proceeds by selecting a parent model and generating a child model through mutations. Two strategies are used: swapping embedding blocks between item groups and selecting blocks for one group. Importantly, only one mutation occurs per round.
    The child model is added to both the seed set and the cache set. Meanwhile, the poorest performing design is removed from the seed set to maintain a constant size. This process of creating child models and updating sets continues for a finite number of rounds.
    At the end of the evolution, the cache set holds the best-performing models, offering a collection of designs with superior performance.

    Another novel perspective views embedding matrix size selection as a regularization problem. Inspired by Soft Threshold Reparameterization~\cite {softThreshold-kusupati2020soft}, different from AMTL, PEP \cite{liu2021pep} directly prunes the embedding matrix $\mathbf{E}$ by learning a threshold.
    Before training the recommendation model, it is argued, according to the Lottery Ticket hypothesis, that the parameters can be initialized by performing an element-wise product between ${\mathbf{E}}_0$ and $m$, where ${\mathbf{E}}_0$ denotes the unpruned embedding parameters, and $m \in \{0,1\}^{N\times D}$ represents the mask matrix generated from $\hat{\mathbf{E}}$.
    
    Another line of research combines learnable anchor embedding matrices to form an optimal embedding matrix. ANT \cite{liang2020ant} and autoDis \cite{guo2021autodis} follow this idea to search for a suitable embedding size. {ANT} first learns a meta embedding $A \in \mathbb{R}^{|A| \times d}$ containing a set of anchor embeddings $A =\{a_1,\cdot\cdot\cdot,a_{|A|}\}, |A|<<|V|$, where $|V|$ is the number of rows in original embedding $E$. Then, it integrates the set of anchor embeddings with a trainable sparse transformation from $A$ to $E$. However, different from the hard selection of anchor embeddings in ANT, autoDis designs a differentiable automatic discretization network to execute a soft selection of meta-embeddings (i.e., anchor embeddings in ANT).

\subsection{Surveys and Future Directions}
    \noindent \textbf{Survey.} 
    In terms of Deep Recommender Systems~(DRS), Zheng et al.~\cite{zheng2022automlsurvey1} and Chen et al.~\cite{Chen2022automlsurvey2} both demonstrate how to perform the autoML technique on each component of DRS, including feature selection, feature embedding, and feature interaction with a different taxonomy of methods.
    \noindent \textbf{Tools.} EasyRec\footnote{https://github.com/alibaba/EasyRec} provides a convenient tool for users to develop and customize the recommendation model. The novel part of EasyRec is that it integrates the autoML API from Alibaba Cloud, which supplies lots of autoML-related services, including auto feature selection and auto feature interaction. 

    \noindent \textbf{Future Direction: Unsupervised NAS in embedding.} Conventionally, the best model architecture is determined based on human-labeled data. However, \cite{liu2020unas} proposed a new unsupervised scheme of NAS called UnNAS in visual tasks. The performance of the sampled architecture needs to be evaluated in an unsupervised manner.
    In addition, the model selected by UnNAS has comparable performance with that chosen by the supervised NAS method. Thus, it is a promising pattern in NAS where we could explore an efficient method.

\section{Quantization}
\label{sec:quantization}

{Quantization, a fundamental technique in information theory, emerges as a promising avenue to bolster the scalability of deep recommender systems. It aims to compress continuous \textit{high-dimensional vectors} into \textit{low-dimensional discrete codes}. In deep recommender systems, there can be tens of thousands of categorical features that are encoded into dense vectors, also known as embeddings. The high memory cost associated with embeddings poses a challenge to the deployment of deep recommender systems.
In this context, quantization provides a strategic pathway to alleviate the memory burden associated with embeddings, thereby enhancing the scalability and operational efficiency of recommender systems.}

\begin{figure}[htbp]
    \centering
    \includegraphics[width=\textwidth]{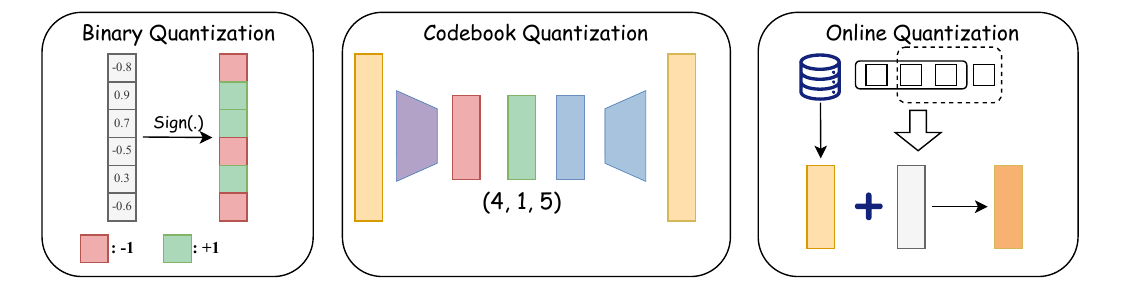}
    \caption{A taxonomy of quantization methods discussed in this survey. 
    (Left) Binary Quantization converts embeddings into binary codes. 
    (Center) Codebook Quantization, such as Product Quantization, 
    compresses vectors into a set of discrete indices from codebooks. 
    (Right) Online Quantization adapts codebooks to streaming data.}
    \label{fig:quantization_methods}
\end{figure}

Quantization compresses the original embedding into a set of codes. The memory cost of this set of codes is much smaller than the original embedding. One can either (1) use such discrete codes as new embeddings or (2) reconstruct the original embedding with a small distortion.
Therefore, quantization allows us to discard the original embedding, leading to pronounced reductions in memory consumption and offering an effective way to address the scalability challenges in recommender systems.
In this section, we classify quantization methods into: (a) binary quantization and (b) codebook quantization based on the form of quantization embedding. In particular, we additionally discuss the application of quantization embedding on online recommender systems. These quantization approaches are summarized in Table~\ref{tab:quantization_final}. Fig.~\ref{fig:quantization_methods} provides a comprehensive overview of these quantization approaches.
\begin{table}[t!]
\centering
\caption{Embedding Quantization}
\label{tab:quantization_final}
\begin{tabular}{>{\raggedright}m{2.5cm}|>{\raggedright}p{5.5cm}|m{6cm}}
\toprule
\multirow{4}{2.5cm}{\raisebox{-8\height}{\textbf{Quantization}}} & 
\textbf{Binary Quantization (\S\ref{ssec:binary_quantization})} & 
\textbf{Key Models:} Discrete Personalized Ranking (DPR)~\cite{zhang2017discrete}, Discrete Deep Learning (DDL)~\cite{zhang2018discrete} \\
\cmidrule(l){2-3}
& 
\multirow{2}{*}{\raisebox{-4\height}{\textbf{Codebook Quantization (\S\ref{ssec:codebook_quantization})}}} & 
\textbf{Unsupervised Methods:} Product Quantization (PQ)~\cite{jegou2010product}, Optimized PQ (OPQ)~\cite{ge2013optimized}, Additive Quantization (AQ)~\cite{babenko2014additive} \\
& & \textbf{Supervised Methods:} Differentiable PQ (DPQ)~\cite{chen2020differentiable}, Product Quantized CF (PQCF)~\cite{lian2020product}, LightRec~\cite{lian2020lightrec}, Distill-VQ~\cite{xiao2022distill}, MoPQ~\cite{xiao2021matching}, xLightFM~\cite{jiang2021xlightfm} \\
\cmidrule(l){2-3}
& 
\textbf{Online Quantization (\S\ref{ssec:online_quantization})} & 
\textbf{Key Models:} Online PQ~\cite{xu2018online}, Online OPQ~\cite{liu2020online}, Online AQ~\cite{liu2021online} \\
\bottomrule
\end{tabular}
\end{table}

\subsection{Binary Quantization}
\label{ssec:binary_quantization}

Binary quantization converts the embedding into a binary code $\mathbf{b} \in\{\pm 1\}^{r}$, where $r$ denotes the length of the binary code. In the binary code representation, the similarity ${x}_{u i}$ between user $u$ and item $i$ can be written as:

\begin{equation}
{x}_{u i}=\frac{1}{2}+\frac{1}{2 r} \mathbf{b}_{u}^{T} \mathbf{d}_{i}
\label{DPR1}
\end{equation}
where $\mathbf{b}_{u}$, $\mathbf{d}_{i}$ denote the binary quantization embedding of user $u$ and item $i$, respectively.

Discrete Personalized Ranking (DPR) \cite{zhang2017discrete} maps embeddings to binary codes by optimizing AUC. However, AUC is typically utilized as an evaluation metric instead of an optimization objective, and optimizing its ranking can be an NP-hard combinatorial optimization problem.
To address this, DPR follows OPAUC~\cite{gao2013one} by approximately optimizing AUC through a least-squares surrogate loss function, instead of directly optimizing AUC rank. The optimization objective of DDL can be formulated as:

\begin{equation}
\min \sum_{(u, i, j) \in D }\frac{1}{|U|\left|I_{u}^{+}\right|\left|I_{u}^{-}\right|}\left(1-\left(\hat{x}_{u i}-\hat{x}_{u j}\right)\right)^{2},
\label{DPR2}
\end{equation}
Where $|I_u^{+}|$ denotes the number of user-item interactions that exist in the dataset, $|I_u^{-}|$ denotes the number of all user-item without interactions, $|U|$ denotes the number of users, and $D=\left\{(u, i, j) |u \in U,i \in I_{u}^{+}, j \in I_{u}^{-} \right\}$. 

In particular, to obtain more compact and high-quality codes, DPR adds the balance constraint and the irrelevant constraint.
The balance constraint encourages maximizing the entropy of the code, and the irrelevant constraint encourages the bits in the code to be as independent as possible.
For this discrete optimization problem, DPR solves it with softening constraints and alternating optimization.
However, DPR does not impose a constraint on the gap between the binary code and the initial embedding, which is improved by Discrete Deep Learning (DDL) \cite{zhang2018discrete}. DDL uses a bag-of-words model to learn the embedding of the text of an item and minimizes the gap between it and the corresponding binary code. 

While binary quantization is theoretically sound and interpretable,  it limits the scalability of quantization embeddings by making them discrete. Compact embeddings may lose vital information. In contrast, codebook quantization is a preferable approach, as it condenses features into a set of codes. Each code can possess multiple potential values. Moreover, codebook quantization fits better with deep learning techniques. It trains continuous codebook elements rather than discrete ones. We'll delve into codebook quantization in the next section.

\subsection{Codebook Quantization}
\label{ssec:codebook_quantization}

Codebook quantization is a family of methods that compress the original embedding matrix into a set of codebooks and indices, allowing for the reconstruction of the embedding used for downstream recommendation tasks. These codebooks and indexes require a small memory footprint, making them suitable for use in various applications. Here, we present three types of codebook quantization methods: unsupervised and supervised codebook quantization.

Most of the existing codebook quantization methods are variations and extensions of Product Quantization (PQ) \cite{jegou2010product}, PQ decomposes the high-dimensional feature space $\mathbb{R}^{D}$ into a cartesian product $\mathcal C={C}^{1} \times \cdots \times \mathcal{C}^{M} \in \mathbb{R}^{M \times K}$ of lower-dimensional subspaces.
So the high-dimensional vector can be compressed into a compositional representation of $M$ codebooks, and the quantization embedding can be obtained by concatenating the $M$ sets of codewords. The objective function of Vanilla PQ can be formulated as follows:
\begin{equation}
\begin{aligned}
&\min _{\mathcal{C}^{1}, \ldots, \mathcal{C}^{M}} \sum_{\mathbf{x}}\|\mathbf{x}-\mathbf{c}(i(\mathbf{x}))\|^{2}, \\
&\text { s.t. } \quad \mathbf{c} \in \mathcal{C}=\mathcal{C}^{1} \times \cdots \times \mathcal{C}^{M}
\end{aligned}
\end{equation}
Where $\mathbf{x}$ denotes the sample features and ${i(\mathbf{x}})$ denotes the quantization encoder that encodes the features into the indexes.

Vanilla PQ does not optimize the objective directly; it applies K-means clustering to each of the $M$ subspaces and compresses each subspace into $K$ codewords, so that each feature vector is compressed into a quantization embedding $i(\mathbf{x})=\{ 1,...,K\}^{M}$. By computing the Euclidean distance, Vanilla PQ can be applied to the nearest neighbor search.
However, Vanilla PQ has a serious problem in that it directly divides the features into different subspaces with equal spacing in an ordered manner and may ignore the possibility of a high correlation between subspaces, which can substantially degrade the performance of quantization. Optimized Product Quantization (OPQ)~\cite{ge2013optimized}
uses the rotation matrix $R \in \mathbb{R}^{M \times M}$ to optimize the decomposition of subspaces to reduce the correlation between subspaces.


The PQ-based approach requires decomposing the embedding space into independent subspaces, so this approach leads to large information loss when the subspace embeddings are strongly correlated. Additive Quantization (AQ) \cite{babenko2014additive} adopts a different strategy from PQ, it directly assigns $M$ codebooks to the whole high-dimensional space, and the $M$ compressed vectors are summed to obtain the quantization embedding.
\begin{equation}
\begin{aligned}
&\min _{\mathcal{C}^{1}, \ldots, \mathcal{C}^{M}} \sum_{\mathbf{x}}\|\mathbf{x}-\sum_{m} \mathbf{c}^{m}(i_{m}(\mathbf{x}))\|^{2} \\
\end{aligned}
\end{equation}
Since optimizing the above equation is a complex combinatorial optimization problem, AQ chooses beam search~\cite{shapiro1992encyclopedia} as the approximation algorithm to select the appropriate codewords for the codebook to complete the quantization.

Both PQ and AQ-based quantization methods work in an unsupervised manner, employing unsupervised k-means clustering or heuristic beam search algorithms to optimize their objective functions. However, these methods fail to address the distortion of embeddings caused by quantization, leading to a partial loss of original embedding information. Consequently, obtaining high-quality embedding becomes challenging.
Differentiable Product Quantization (DPQ) \cite{chen2020differentiable} 
proposes softmax-based and centroid-based methods to minimize reconstruction loss approximately between raw embeddings and quantization embeddings. This approach consistently leads to notably higher-quality quantization embeddings compared to traditional unsupervised PQ algorithms.

Supervised methods \cite{chen2020differentiable,yue2016deep} enhance quantization via reconstruction loss, but recent research \cite{zhan2021jointly,xiao2021matching,lian2020lightrec} shows it's insufficient. Tailoring objectives for specific tasks in recommender systems proves more effective.
Product Quantized Collaborative Filtering (PQCF) \cite{lian2020product} challenges separate user and item quantization using PQ. Misaligned coordinates make this suboptimal. PQCF minimizes rating prediction loss, moving from Euclidean to the inner product space. It rotates ${\mathbf C_{U}}$ and ${\mathbf C_{V}}$ with orthogonal matrix $\mathbf{H}$, aligning spaces and addressing PQ-based methods issues.
Alongside optimizing downstream recommendations, new optimization goals have emerged. Distill-VQ \cite{xiao2022distill}, inspired by knowledge distillation, sets its objective as a similarity function measuring differences in relevance score distributions between teacher and student models (e.g., KL divergence). 
Another innovation, LightRec \cite{lian2020lightrec}, enhances reconstruction loss with two functions, minimizing differences in user-item ratings pre- and post-quantization, as well as alterations in recommendation ranking. Matching-oriented Product Quantization (MoPQ) \cite{xiao2021matching} shows that better quantization reconstruction doesn't always mean better downstream performance. MoPQ improves accuracy through contrastive learning, modeling query-quantization matching via multinoulli process.
\subsection{Online Quantization}
\label{ssec:online_quantization}
Modern recommender systems face constant influxes of new users, and traditional quantization methods (PQ, AQ) lack the capacity to manage streaming data, thus presenting challenges in real-world applications. Similarly, prior online hash algorithms require recalculating all user embeddings upon new user arrivals, which is often impractical in recommendation systems.
To tackle this, Online PQ \cite{xu2018online} posits that the impact of new data on the codebook is minimal due to the smaller size of incremental data. It updates only codewords for new data points without altering quantization embedding for original data.
Online PQ maintains a sliding window for data processing, continuously applying K-means clustering for new codewords. However, being unsupervised, Online PQ doesn't optimize quantization error, risking information loss and low-quality embedding. To address this, Online OPQ \cite{liu2020online} extends to streaming data by solving the orthogonal procrustes problem to ensure subspace orthogonality during codebook updates. Online AQ \cite{liu2021online} extends AQ, maintaining consistent objective functions. For streaming data adaptability, Online AQ derives codebook update strategies and related regret bounds via linear regression closed solutions and matrix inversion lemma~\cite{tylavsky1986generalization}. Unlike AQ's beam search for codeword selection, Online AQ introduces an efficient block beam search, a trade-off between hill climbing and beam search.

\subsection{Surveys and Future Directions}
\noindent \textbf{Survey.} To the best of our knowledge, there does not exist any survey on quantization embedding for recommender systems. In addition, existing surveys on quantization either focus on image coding \cite{nasrabadi1988image} or are limited to product quantization \cite{ali2005quantization}.
Some existing reviews on codebooks also belong to the scope of quantization. For example, \cite{lu2010survey} discusses a variety of VQ Codebook Generation algorithms, and \cite{ramanan2012review} discusses codebook design in the field of object recognition.

\noindent \textbf{Future Direction 1: Multi-task Quantization.} Previous studies~\cite{zhan2021jointly,xiao2021matching,lian2020lightrec} have highlighted that optimizing solely for reconstruction loss is inadequate for producing high-quality quantization embeddings. For better quantization, tailoring learning tasks to specific downstream recommendation contexts can enhance pattern capture. Introducing multi-task learning objectives alongside quantization distortion minimization as a new objective empowers models to grasp more valuable patterns, potentially leading to improved generalization. 

\noindent \textbf{Future Direction 2: Dynamic Codebook Quantization.}
In online recommender systems, accommodating changing user preferences is crucial. Accurate quantization is essential for domain features that exert a significant impact on user preferences, like user interaction features. However, less impactful domain features like user location may need smaller codebooks. 
Exploring methods for capturing changes in user preferences and dynamically adjusting the codebook size for each domain feature accordingly is a promising research direction that warrants attention. Such in-depth research could considerably enhance the operational efficiency and overall effectiveness of online recommender systems.

\section{LLM-Driven Embedding Enhancement}
\label{sec:llm_embedding}
Recent years have witnessed remarkable advancements in the field of large models, particularly Large Language Models (LLMs), which have demonstrated transformative potential across various domains. These models, with their strong language understanding and reasoning capabilities, are now increasingly being leveraged within recommender systems to overcome the limitations of traditional embedding techniques and further elevate the quality of recommendations. This chapter introduces the paradigm of "Large Model Enhanced Modeling" in recommender systems, where the power of these models is harnessed to refine and augment the embedding layer, leading to significant improvements in capturing user preferences and item characteristics. This section will explore three key areas where large models are making a substantial impact on embedding techniques: \textbf{(1) LLM embedding as the direct semantic supplement}, and \textbf{(2) LLM embedding as guidance}. By delving into these aspects, this section aims to provide a comprehensive understanding of how large models complement and extend the capabilities of existing centralized embedding methods discussed in this survey, including matrix factorization, sequential modeling, and graph embeddings, as well as techniques addressing scalability like AutoML, hashing, and quantization. Rather than replacing these established methods, large models offer a powerful means of enhancing their effectiveness. An overview of these enhancement strategies is provided in Table~\ref{tab:llm_embedding_final}.
\begin{table}[t!]
\centering
\caption{LLM-Driven Embedding Enhancement}
\label{tab:llm_embedding_final}
\begin{tabular}{>{\raggedright}m{2.5cm}|>{\raggedright}p{5.5cm}|m{6cm}}
\toprule
\multirow{5}{2.5cm}{\raisebox{-3\height}{\textbf{\begin{tabular}{@{}l@{}}LLM-Driven \\ Embedding \\ Enhancement\end{tabular}}}} & 
\multirow{3}{5.5cm}{\raisebox{-3\height}{\textbf{\begin{tabular}{@{}l@{}}LLM Embedding as \\ Semantic Supplement (\S\ref{ssec:llm_supplement})\end{tabular}}}} & 
\textbf{Direct Inference:} ChatRec~\cite{gao2023chat}, GeneRec~\cite{wang2023generative}, TedRec~\cite{xu2024sequence}, LRD~\cite{yang2024sequential}, SemSR~\cite{narwariya2025semsr} \\
& & \textbf{With Trainable Modules:} LLM-ESR~\cite{liu2024llm}, AlphaRec~\cite{sheng2024language}, Laser~\cite{lin2025large}, SRA-CL~\cite{cui2025semantic}, SAID~\cite{hu2024enhancing} \\
& & \textbf{Fine-tuning LLM/Embeddings:} PEPLER~\cite{li2023personalized}, PPR~\cite{wu2022personalized}, LLM-CF~\cite{sun2024large}, TPAD~\cite{su2025distilling}, LLMEmb~\cite{liu2025llmemb}, SeRALM~\cite{ren2024enhancing}, LLM2Rec~\cite{he2025llm2rec} \\
\cmidrule(l){2-3}
& 
\multirow{2}{5.5cm}{\raisebox{-2\height}{\textbf{\begin{tabular}{@{}l@{}}LLM Embedding as Guidance (\S\ref{ssec:llm_guidance})\end{tabular}}}} & 
\textbf{Self-Guidance:} LLM4SBR~\cite{qiao2024llm4sbr}, DaRec~\cite{yang2025darec} \\
& & \textbf{Cross-Embedding Guidance:} Jia et al.~\cite{jia2025improving}, PAD~\cite{wang2025pre}, LLM4CDSR~\cite{liu2025bridge}, EIMF~\cite{qiao2024llm}, LGMcRec~\cite{meng2025lgmcrec} \\
\bottomrule
\end{tabular}
\end{table}
\subsection{LLM Embedding as Semantic Supplement}
\label{ssec:llm_supplement}
\begin{figure}[htbp]
\centering
\includegraphics[scale=0.6]{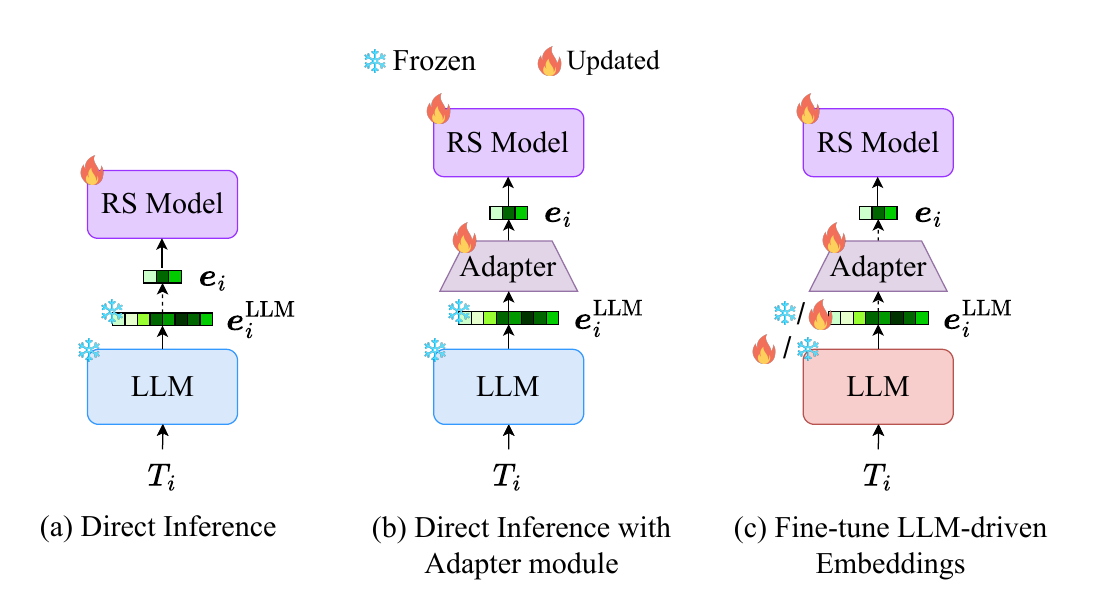}
\caption{Categories of LLM Embeddings as Semantic Supplement. (a) Generate embeddings using frozen LLMs and convey them to recommendation models after dimensional reduction. (b) Employ a trainable adapter to relieve the information loss of LLM embeddings. (c) Fine-tune LLMs or the embeddings to enhance the semantic information for recommendation tasks.}
\label{llmemb}
\end{figure}

Natural languages contain various information about users and items, but it is hard to effectively add textual features into recommender systems. Recently, there has been a notable surge of interest in the incorporation of Large Language Model embeddings into recommendation systems to supplement semantic information~\cite{wu2023survey, liu2025large,xiao2025mars}. In the recommendation task, given the interaction sequence of a given user $u$ can be denoted as $\mathcal{\boldsymbol{S}}^{(u)}=[v_1^{(u)}, v_2^{(u)}, \dots, v_{\vert \mathcal{S}^{(u)} \vert}^{(u)}]$, the aim is to predict the next item the users would like to select, which can be formulated as:
\begin{equation}
    \arg\max_{v_j\in\mathcal{V}} P(v_{N+1}=v_j \vert \mathcal{\boldsymbol{S}}),
\end{equation}
where $v_j\in\mathcal{V}$ is the selected item in the item set, $N$ is the number of items that the user interacts. In the LLM-driven embedding enhancement strategies, the embedding procedure of the item sequences is necessary, which can be written as $\text{LLM}_{emb}(\cdot)$. This function usually denotes that the last hidden state of the LLM is regarded as the embedding of a given item sequence. However, the item ids are labels that LLMs are hard to understand, so a prompt $T_i=[I, A_1, A_2, \dots, A_K]$ should also be constructed for the $i$-th item, where $I$ is the instruction for LLM generation, and $A_k (k\in [1, K])$ is the attribute of the item. Then the LLM-driven embedding $\boldsymbol{e}_i^{\text{LLM}}$ can be generated:
\begin{equation}
    \boldsymbol{e}_i^{\text{LLM}} = \text{LLM}(T_i), \ \ i\in[1,  N].
\end{equation}
As shown in Figure \ref{llmemb}, these LLM embeddings can be utilized for \textbf{direct inference}, and they can be \textbf{fine-tuned} for capturing more fine-grained semantic relationships. In this subsection, we will discuss the two categories in detail.

\noindent \textbf{Direct Inference of Embeddings Utilizing LLMs.} 
In this strategy, using LLMs, we can extract embedding vectors
from textual descriptions of users, items, and user-item interactions. This is particularly effective in scenarios with
limited data for user-item interactions. For example, ChatRec ~\cite{gao2023chat} generates embeddings through prompting a LLM with prompts generated by a multi-input prompt constructor module. The module forms a natural language paragraph that
captures user intent and item details by considering various inputs, such as user-item history interactions, user profiles,
specific user queries, and dialog history. GeneRec ~\cite{wang2023generative} uses user profiles, historical feedback, and collected Web data to create personalized embeddings through LLMs. Specifically, it can generate personalized embeddings by analyzing user instructions and feedback, resulting in tailored content like landscape micro-videos in the chosen style.
TedRec ~\cite{xu2024sequence} and LRD ~\cite{yang2024sequential} utilize frozen language models to directly generate representations for embeddings of semantic information and IDs. SemSR ~\cite{narwariya2025semsr} generates the semantics embedding from the LLMs and integrates it into a trainable end-to-end Semantic sequential recommender.

However, directly integrating such embeddings might lead to information loss ~\cite{liu2025large}. Therefore, recently, more research has focused on enhancing the embeddings from LLMs by adding trainable modules to tailor the embeddings to LLMs more effectively. One representative work is LLM-ESR ~\cite{liu2024llm}, where a trainable adapter module is designed for effective original semantics retention in both the trainable collaborative embedding layer and trainable cross attention. In AlphaRec ~\cite{sheng2024language}, an MLP serves as a bridge between the language representations and ID representations. SRA-CL ~\cite{cui2025semantic} utilizes the sequence encoder for user embeddings from LLMs to enhance the information representation capability of contrastive samples. Laser ~\cite{lin2025large} employs an adapter to enhance features by leveraging the user and item embeddings derived from the LLM. In addition, trainable encoder modules can also be applied to generating global interaction embeddings and learning user preferences in the cross-domain recommendation ~\cite{liu2025bridge}. SAID ~\cite{hu2024enhancing} semantically aligned the item embeddings for sequential recommenders using a projector to derive an embedding vector, which is utilized to preserve fine-grained semantic information. Uniquely, the projector is aligned within the LLM for better embedding generation, followed by the training together with the recommenders.

\noindent \textbf{Fine-tuning LLM-Driven Embeddings.} Embeddings can be enhanced not only by adding a trainable module to learn more effective representations, but also by fine-tuning embedding vectors directly or by fine-tuning LLM itself to adapt the embedding generation, thereby enhancing the quality of embeddings and boosting the overall performance of the system. 
For example, in the case of PEPLER~\cite{li2023personalized}, it successfully embedded IDs into a semantic space for better embedding explanation ability. PEPLER utilizes a two-stage approach for fine-tuning embedding vectors and model parameters. 
PPR~\cite{wu2022personalized} learns personalized embedding vectors by employing prompt-level augmentation, which involves randomly masking elements in user embedding vectors, and behavior-level augmentation that masks some items in the user-item interaction sequence. 
LLM-CF ~\cite{sun2024large} designs a data mixture method to distill the world knowledge of LLMs into recommenders, also enhancing the representation ability of item embeddings. 

Since the LLMs are the key source of embeddings' semantic information, fine-tuning LLMs can also be effective in conveying more knowledge to the embeddings. For example, TPAD ~\cite{su2025distilling} fine-tunes the multimodal LLMs to generate better knowledge item embeddings, thus aligning both transitional patterns and knowledge patterns in one unified LLM more effectively. LLMEmb ~\cite{liu2025llmemb} employs a supervised contrastive fine-tuning approach to train the LLM together with the adapter module to learn effective information from LLM embeddings. Laser ~\cite{zhang2024laser} uses a bi-tuning method by inserting trainable virtual prefix and suffix tokens into the input sequence. These semantic tokens enhance the LLMs' capability of generating high-quality embeddings. SeRALM ~\cite{ren2024enhancing} applies an alignment training to refine LLMs’ embedding generation according to the feedback from the sequential recommenders, achieving a cycling optimization paradigm. Moreover, the fine-tuning of embeddings can be combined with fine-tuning of LLMs in one framework to collaboratively learn better embedding utilization. For instance, LLM2Rec ~\cite{he2025llm2rec} employs an innovative two-stage training framework that uses a trainable LLM to generate embedding and inherently enhances both semantic information and collaborative filtering signals rather than merely using the LLM as a feature extractor or training guide.

\subsection{LLM Emebdding as Guidance}
\label{ssec:llm_guidance}
Unlike approaches that directly utilize LLM embeddings as input features, this enhancing strategy treats embeddings not as primary representations, but rather as high-level semantic guidance to steer the training process or facilitate parameter synthesis ~\cite{liu2025large}. By decoupling representation learning from downstream model adaptation, these methods aim to transfer and guide the learning process, as well as enhance the embedding's expressive capability in recommendation models. The guiding mechanism of the LLM embeddings can be categorized into \textbf{Self-Guidance} and \textbf{Cross-Embedding Guidance}, which will be introduced in detail in this subsection.

\noindent \textbf{Self-Guidance.} LLM embeddings can be used as a guide to identify representative items of users. LLM4SBR ~\cite{qiao2024llm4sbr} and LLM-ESR ~\cite{liu2024llm} enhance the performance of recommendation models by utilizing LLM embeddings to identify user preferences and similar user behaviors, thereby refining the modeling of diverse behavior patterns and guiding the learning process of embedding utilization in recommendation models. DaRec ~\cite{yang2025darec} proposes a plug-and-play disentanglement framework that separates LLM and collaborative embeddings into shared and task-specific components, then aligns the shared representations. LLMEmb ~\cite{liu2025llmemb} employs a supervised contrastive fine-tuning approach to facilitate self-guided training for the effective utilization of LLM embeddings, thereby alleviating the long-tail problems in recommendation.

\noindent \textbf{Cross-Embedding Guidance}. Basic embeddings from users and items, and the LLM embeddings of users and items, share potential semantic relationships. In addition, they contribute different semantic information for recommenders. Therefore, the LLM embeddings can also serve as guidance for learning basic embeddings. Jia et al. ~\cite{jia2025improving} demonstrate a typical cross-embedding guidance paradigm, where the LLM embedding of users and items can be applied to refine the corresponding basic embeddings. PAD ~\cite{wang2025pre} constructs LLM embeddings of item titles and aligns them with the initial item ID embeddings. LLM4CDSR ~\cite{liu2025bridge} employs the global interaction embeddings from LLMs and hierarchical LLMs profile embeddings to guide the learning of local user embeddings and item embeddings. EIMF ~\cite{qiao2024llm} constructs the embeddings of clustered user behaviors, which are also used to guide the learning of basic behavior embeddings. LGMcRec ~\cite{meng2025lgmcrec} uses the LLM to generate user and item profile embeddings, which are applied to enhance the embeddings of user and item interaction embeddings learned from graph neural networks.

\subsection{Surveys and Future Directions}
\noindent \textbf{Survey:} LLMs can play an important role in recommendation systems by extracting features and deriving high-quality embeddings through the incorporation of the rich semantic information embedded in the pretrained or fine-tuned LLMs. These embeddings can be enhanced through the specially designed and trained adapter modules or fine-tuned methods on the embeddings. LLM-enhanced embeddings have demonstrated their potential in enhancing performance in applications like video recommendation~\cite{liu2023llmrec} and recommendation explanation generation~\cite{zhou2023gpt}.

\noindent \textbf{Future Direction 1: Computational Efficiency and Deployment.} While LLMs show great promise for enhancing recommendation systems through embedding enhancement, the computational demands pose significant implementation barriers~\cite{wu2023survey}. These models require substantial computational resources for inference and fine-tuning, making them challenging to deploy in real-world recommendation scenarios where millisecond-level response times are often required~\cite{zhang2025glint, zhang2024dns}. This challenge necessitates research into efficient embedding caching strategies, model compression techniques, and lightweight adapter architectures specifically optimized for recommendation contexts. Integration challenges also present significant hurdles in determining optimal strategies for combining LLM-derived semantic knowledge with traditional collaborative filtering signals, maintaining high recommendation quality while reducing computational overhead, and striking the right balance between personalization and generalization when utilizing LLM embeddings~\cite{wu2023survey}. Scalability remains a critical concern, particularly in handling large-scale user-item matrices and efficiently updating LLM-enhanced embeddings as new items and users are added to the system~\cite{zhang2025glint}.

\noindent \textbf{Future Direction 2: Fairness and Privacy Enhancement.} Another crucial concern relates to bias and fairness in LLM-enhanced recommendations. LLMs can inherit and amplify societal biases present in their training data, including gender, racial, and other demographic biases, which may lead to unfair or discriminatory recommendations~\cite{liu2023trustworthy, thakur2023unveiling}. This calls for dedicated research into methods for detecting and mitigating bias in LLM-enhanced embeddings, developing fairness-aware fine-tuning approaches, and establishing transparent evaluation metrics for assessing recommendation fairness. Privacy considerations in processing user data through LLMs add another layer of complexity to these integration challenges~\cite{liu2023trustworthy}. Developing effective techniques for addressing cold-start problems while leveraging LLM knowledge is also essential. These challenges collectively present important research opportunities for making LLM-enhanced recommendation systems more practical and trustworthy for real-world applications.

\section{Conclusion}

In summary, the use of embeddings in recommender systems has shown great promise in recent years, demonstrating significant advantages in capturing complex relationships between users and items. From traditional matrix factorization approaches to advanced sequential modeling and graph-based methods, embedding techniques have evolved to address various challenges in recommendation scenarios. The integration of self-supervised learning has further enhanced embedding quality without requiring extensive labeled data, while practical considerations have led to innovations in embedding optimization through AutoML, hashing, and quantization techniques. The recent emergence of large language models has opened new possibilities for enhancing embedding quality through semantic understanding and knowledge transfer, though challenges remain regarding computational efficiency and practical deployment. However, there are still several critical challenges to overcome, including scalability issues, the need for better interpretability, and concerns about fairness and bias in learned representations. This survey provides a comprehensive overview of embedding techniques in recommender systems, covering their theoretical foundations, practical applications, and effectiveness across different scenarios. By systematically examining current methodologies and highlighting future research directions, this work aims to serve as a valuable resource for researchers and practitioners in the field. As recommender systems continue to evolve and tackle more complex scenarios, embedding techniques will remain crucial for their success, and the insights shared in this survey pave the way for future innovations and improvements in this rapidly developing landscape.

\section*{Acknowledge}
This research was partially supported by National Natural Science Foundation of China (No.62502404), Hong Kong Research Grants Council's Research Impact Fund (No.R1015-23), Research Grants Council's Collaborative Research Fund (No.C1043-24GF), Research Grants Council's General Research Fund (No.11218325), Institute of Digital Medicine of City University of Hong Kong (No.9229503), Huawei (Huawei Innovation Research Program), Tencent (CCF-Tencent Open Fund, Tencent Rhino-Bird Focused Research Program), Alibaba (CCF-Alimama Tech Kangaroo Fund No. 2024002), Ant Group (CCF-Ant Research Fund), Didi (CCF-Didi Gaia Scholars Research Fund), Kuaishou, and Bytedance.

\bibliographystyle{ACM-Reference-Format}
\bibliography{bibtex/bib/hash,bibtex/bib/CF&MF,bibtex/bib/quan,bibtex/bib/autoML,bibtex/bib/graph,bibtex/bib/self-learning,bibtex/bib/application,bibtex/bib/intro,bibtex/bib/LLM}










\end{document}